\documentclass{aa}  

\usepackage{graphicx}
\usepackage{txfonts}
\usepackage{amsmath}	
\usepackage{xcolor,colortbl}
\usepackage{soul}
\usepackage{xspace}
\usepackage{enumerate}
\usepackage{placeins}
\usepackage{makecell}
\usepackage[caption=false]{subfig}
\usepackage{pdflscape}
\usepackage{rotfloat}
\usepackage{adjustbox}
\usepackage{booktabs}
\usepackage{makecell}
\usepackage{multirow}
\usepackage{natbib}
\bibpunct{(}{)}{;}{a}{}{,}
\usepackage[pdftex,bookmarks=true,colorlinks=true,linkcolor=blue,citecolor=blue, urlcolor=blue,breaklinks]{hyperref}      
\usepackage{scalerel}
\usepackage{pgfplots}
\usepackage{tikz}

\usetikzlibrary{svg.path}
\usetikzlibrary{shapes.geometric}

\definecolor{orcidlogocol}{HTML}{A6CE39}
\tikzset{
  orcidlogo/.pic={
    \fill[orcidlogocol] svg{M256,128c0,70.7-57.3,128-128,128C57.3,256,0,198.7,0,128C0,57.3,57.3,0,128,0C198.7,0,256,57.3,256,128z};
    \fill[white] svg{M86.3,186.2H70.9V79.1h15.4v48.4V186.2z}
                 svg{M108.9,79.1h41.6c39.6,0,57,28.3,57,53.6c0,27.5-21.5,53.6-56.8,53.6h-41.8V79.1z M124.3,172.4h24.5c34.9,0,42.9-26.5,42.9-39.7c0-21.5-13.7-39.7-43.7-39.7h-23.7V172.4z}
                 svg{M88.7,56.8c0,5.5-4.5,10.1-10.1,10.1c-5.6,0-10.1-4.6-10.1-10.1c0-5.6,4.5-10.1,10.1-10.1C84.2,46.7,88.7,51.3,88.7,56.8z};
  }
}

\newcommand\orcidicon[1]{\href{https://orcid.org/#1}{\mbox{\scalerel*{
\begin{tikzpicture}[yscale=-1,transform shape]
\pic{orcidlogo};
\end{tikzpicture}
}{|}}}}
\bibpunct{(}{)}{;}{a}{}{,}            

\pgfplotsset{compat=1.17} 

\definecolor{scc}{rgb}{0.54, 0.17, 0.89}

\definecolor{gri}{rgb}{1.0, 0.16, 0.64}

\definecolor{yle}{rgb}{1.0, 0.16, 0.64}

\definecolor{cadmiumred}{rgb}{0.89, 0.0, 0.13}

\definecolor{darkpastelgreen}{rgb}{0.01, 0.75, 0.24}

\definecolor{lavender}{rgb}{0.9, 0.9, 0.98}

\definecolor{dem}{rgb}{0.858, 0.188, 0.478}

\definecolor{inferno0}{rgb}{0.002,0.004,0.329}
\definecolor{inferno1}{rgb}{0.144,0.018,0.499}
\definecolor{inferno2}{rgb}{0.311,0.063,0.640}
\definecolor{inferno3}{rgb}{0.477,0.125,0.737}
\definecolor{inferno4}{rgb}{0.647,0.197,0.764}
\definecolor{inferno5}{rgb}{0.811,0.290,0.706}
\definecolor{inferno6}{rgb}{0.945,0.411,0.553}
\definecolor{inferno7}{rgb}{0.993,0.588,0.389}
\definecolor{inferno8}{rgb}{0.996,0.788,0.217}
\definecolor{inferno9}{rgb}{0.987,0.998,0.645}

\definecolor{viridis0}{rgb}{0.267,0.005,0.329}
\definecolor{viridis1}{rgb}{0.283,0.141,0.458}
\definecolor{viridis2}{rgb}{0.254,0.265,0.530}
\definecolor{viridis3}{rgb}{0.207,0.372,0.553}
\definecolor{viridis4}{rgb}{0.164,0.471,0.558}
\definecolor{viridis5}{rgb}{0.128,0.566,0.551}
\definecolor{viridis6}{rgb}{0.157,0.668,0.508}
\definecolor{viridis7}{rgb}{0.369,0.788,0.382}
\definecolor{viridis8}{rgb}{0.678,0.865,0.183}
\definecolor{viridis9}{rgb}{0.993,0.906,0.144}

\newcommand{\ViridisColorbar}[1]{
    \begin{tikzpicture}
        \foreach \i in {0,...,9} {
            \fill[viridis\i] (\i * #1, 0) rectangle ({(\i + 1) * #1}, 0.5);
        }
        
        \draw[black] (0, 0) rectangle ({10 * #1}, 0.5);

    \end{tikzpicture}
}

\makeatletter
\renewcommand*\aa@pageof{, page \thepage{} of \pageref*{LastPage}}
\makeatother

\begin{document} 
    \title{Navigating AGN variability with self-organizing maps}

      \titlerunning{Navigating AGN Variability with self-organizing maps}
    \authorrunning{Y. Maruccia et al.}

   \author{Ylenia~Maruccia \inst{1,\orcidicon{0000-0003-1975-6310}} \and
           Demetra~De~Cicco \inst{2,1,3\orcidicon{0000-0001-7208-5101}} \and
           Stefano~Cavuoti \inst{1,4,\orcidicon{0000-0002-3787-4196}} \and \\
           Giuseppe~Riccio \inst{1,\orcidicon{0000-0001-7020-1172}} \and 
           Paula Sánchez-Sáez\inst{5,3,\orcidicon{0000-0003-0820-4692}} \and
           Maurizio~Paolillo \inst{2,1,4,\orcidicon{0000-0003-4210-7693}} \and \\
           Noemi~Lery~Borrelli \inst{2} \and
           Riccardo~Crupi \inst{6,\orcidicon{0009-0005-6714-5161}} \and
           Massimo~Brescia \inst{2,1,\orcidicon{0000-0001-9506-5680}}
          }

   \institute{INAF - Astronomical Observatory of Capodimonte, Via Moiariello 16, I-80131 Napoli, Italy
        \and
              Department of Physics ``E. Pancini'', University Federico II of Napoli, Via Cinthia 21, I-80126 Napoli, Italy
        \and
                    Millennium Institute of Astrophysics (MAS), Nuncio Monse\~nor Sotero Sanz 100, Providencia, Santiago, Chile  
        \and
              INFN section of Naples, via Cinthia 6, I-80126, Napoli, Italy    
        \and
        European Southern Observatory, Karl-Schwarzschild-Strasse 2, 85748 Garching bei M\"unchen, Germany 
         \and
            Intesa Sanpaolo S.p.A., Corso Inghilterra 3, I-10138, Turin, Italy
             }

   \date{Received Month XX, YYYY; accepted Month XX, YYYY}

  \abstract
   {The classification of active galactic nuclei (AGNs) is a challenge in astrophysics. Variability features extracted from light curves offer a promising avenue for distinguishing AGNs and their subclasses. This approach would be very valuable in sight of the Vera C. Rubin Observatory Legacy Survey of Space and Time (LSST). }
   {Our goal is to utilize self-organizing maps (SOMs) to classify AGNs based on variability features and investigate how the use of different subsets of features impacts the purity and completeness of the resulting classifications.}
   {We derived a set of variability features from light curves, similar to those employed in previous studies, and applied SOMs to explore the distribution of AGNs subclasses. We conducted a comparative analysis of the classifications obtained with different subsets of features, focusing on the ability to identify different AGNs types.}
   {Our analysis demonstrates that using SOMs with variability features yields a relatively pure AGNs sample, though completeness remains a challenge. In particular, Type 2 AGNs are the hardest to identify, as can be expected. These results represent a promising step toward the development of tools that may support AGNs selection in future large-scale surveys such as LSST.}
   {}

   \keywords{Galaxies: active  -- Methods: data analysis     -- Methods: statistical
          }

   \maketitle

\section{Introduction}
The field of time-domain astronomy is next to the beginning of a revolutionary era, 
driven by the advent of a new generation of telescopes designed for wide, deep, and high-cadence sky surveys. This revolution will be led by the Legacy Survey of Space and Time (LSST; see, e.g., \citealt{lsst,ivezic19}). The LSST, to be conducted with the Simonyi Survey Telescope at the Vera C. Rubin Observatory, promises to dramatically enhance our understanding of active galactic nuclei (AGNs) in several key areas, such as the demography of AGNs, their luminosity function, their evolution, and the role they play in shaping their host galaxies \citep[e.g.,][]{brandt2018active,raiteri2018blazars}. Hopefully this will provide new insights into the physics and the structure behind the AGNs formation and evolution.

The primary LSST survey, known as the Wide-Fast-Deep (WFD) survey, will focus on an area of approximately 19.6k square degrees. This vast region is expected to be surveyed around 800 times over a decade, utilizing a large amount of the available observing time. Beside that a portion of the time will be dedicated to ultra-deep surveys of well-known areas, collectively referred to as deep drilling fields (DDFs; e.g., \citealt{brandt2018active,scolnic}). These DDFs are regions where extensive multiwavelength information is already available from previous surveys.

The initial 10-year observing program includes a proposal for high-cadence (up to ${\sim}14,000$ visits) multiwavelength observations of an area of approximately 9.6 square degrees per DDF. These observations aim to reach impressive coadded depths of ${\sim}28.5$ mag in the \emph{ugri} bands, ${\sim}28$ mag in the \emph{z} band, and ${\sim}27.5$ mag in the \emph{y} band. Given these characteristics, the DDFs will serve as excellent laboratories for AGNs science.

The advent of wide-field surveys like LSST will push time domain astronomy in an era of unprecedented data volume, with information about millions of sources being collected nightly. This data deluge highlights the urgent need for scalable, automated, and effective methods to analyze sources, identify candidates, and characterize their physical properties. 

To tackle similar challenges related to large and complex datasets, several other disciplines, such as healthcare \citep{Tangaro2015,pei2023financial}, tourism \citep{solazzo2022exploit}, the banking sector \citep[][]{maruccia2025anomaly}, financial trading \citep{jaiswal2023breast}, and environmental monitoring \citep{licen2023self}, have increasingly adopted machine learning (ML) algorithms. Astronomy is now following the same trend, with ML techniques becoming widely used across a variety of applications \citep[e.g.,][]{Masters2015,Djorgovski_2016,cavuoti2017metaphor,Mahabal2017, d2018return,Baron2019, doorenbos2022ulisse,Soo2023,Angora2023,cavuoti24}. Many of these ML algorithms rely on supervised training, utilizing ``labeled'' sets (LSs) of data. These LSs consist of samples of objects with known classifications, characterized by a set of features selected based on the properties of interest. The efficacy of the training process is heavily dependent on the chosen features and the balance of the selected LS, which should ideally provide the most complete and unbiased sampling possible of the source population to be studied, through broad and homogeneous coverage of the parameter space (though in practice this ideal is never fully achieved, often not even partially). 
In light of these limitations, the exploration of unsupervised methods (i.e., approaches that do not rely on a priori knowledge from LSs) has gained increasing interest in the astronomical community \citep[e.g.,][and literature inside]{fotopoulou2024review}. These techniques offer a complementary strategy to supervised learning, allowing for the identification of novel patterns, groups, or anomalies in complex datasets. For these reasons, exploring unsupervised methods represents a valuable and necessary complement to supervised approaches that is worth being accomplished.

Among the current selection of DDFs is the Cosmic Evolution Survey (COSMOS; \citealt{scoville07b}) field, renowned as one of the most thoroughly studied extragalactic survey regions in the sky. 
The present work is part of a series \citep[e.g.,][]{decicco15,decicco19,decicco21} focused on the identification of AGNs in a 1 square degree area in the COSMOS field making use of data from the VLT Survey Telescope (VST; \citealt{VST}). The selection is based on optical variability and the series of works overall also aims to assess the performance of variability selection in the DDFs. Variability is known to characterize AGNs at all wavelengths, with timescales and amplitudes depending on the observing waveband. Specifically, the VST-COSMOS dataset consists of 54 \emph{r}-band visits from the SUpernova Diversity And Rate Evolution (SUDARE; \citealt{botticella13}) survey, originally designed to detect and characterize supernovae \citep{cappellaro15,botticella17}, and covering a 3.3 year baseline. 

These characteristics allow us to explore regions of the variability parameter space that have been poorly investigated to date. Indeed, the VST-COSMOS dataset stands out as one of the few that combine both a considerable depth and a high observing cadence, two fundamental requirements for variability surveys that are often mutually exclusive in ground-based observations. For such reasons this dataset represents a benchmark in order to develop and test tools that will be then applied to the LSST data. \cite{decicco21}, in particular, made use of statistical derived parameters in order to train a random forest classifier \citep{breiman2001random}.

As was mentioned before, the present work, is part of a series of studies in which our team has explored various approaches to AGNs identification, ranging from classical variability analysis (see \cite{decicco15,decicco19} to supervised ML methods (see \cite{decicco21,decicco2025}). In this work, we decided to investigate an unsupervised method to assess whether it could improve the results, at least for a specific subclass of AGNs. For this purpose, by using the same dataset and the extracted features as in \cite{decicco21}, we applied a self-organizing map (SOM, \citealt{kohonen2001basic}) in order to verify if it is possible to effectively separate different classes of astrophysical sources (AGNs, stars, and galaxies), without directly using the labels, which may be incomplete or biased. Labels were, however, utilized in order to interpret and validate the results of the network. 
SOMs can project high-dimensional data (such as light curve features and colors) onto a two-dimensional map while preserving topological relationships. This capability is particularly useful to visualize and somehow identify regions in which similar objects are falling. 
SOMs have been successfully employed in several astrophysical applications, including applications on stellar and galaxy spectra \citep[e.g.,][]{teimoorinia2022mapping}, classification of images \citep[e.g.,][]{gupta2022discovery,holwerda2022galaxy}, estimation of physical parameters of galaxies \citep{hemmati2019bringing,davidzon2022cosmos2020}, demonstrating their potential as a powerful tool for exploring complex and high-dimensional datasets.

In this perspective, we further explore the potential of the SOM to identify unlabeled sources with similar properties by leveraging their proximity to well-defined prototypes in the map, thus demonstrating its usefulness as a tool for unsupervised classification and label propagation in the presence of incomplete or uncertain labels.

The paper is organized as follows. In Sec.~\ref{sec:data} we describe the data used for this work, while in Sec.~\ref{sec:method} we present the methodology applied. In Sec.~\ref{sec:results} we discuss the results, and finally in Sec.~\ref{sec:conclusion} we draw our conclusions.   

\section{Data}\label{sec:data}

For our study, we utilized the same dataset as employed in several previous works (see for instance \citealt{decicco19,decicco21,decicco22} and \citealt{cavuoti24}). This dataset comprises $r$-band observations of the COSMOS field, obtained using the VST over three observing seasons from December 2011 to March 2015. These observations are part of a long-term initiative to monitor the LSST Deep Drilling Fields prior to the commencement of Vera Rubin Telescope operations.

The VST, a 2.6-meter optical telescope, covers a Field of View (FoV) of 1 square degree with a single pointing (pixel scale: $0.214$\arcsec/pixel). The dataset encompasses a total of 54 visits across three observing seasons, with two intervening gaps. 

The $r$-band observations were originally designed with a three-day observing cadence, although the actual cadence was subject to observational constraints. The single-visit depth reaches $r \lesssim 24.6$ mag for point sources, at a ${\sim}5\sigma$ confidence level; this is comparable to the single-visit depth of $r \sim 24.7$ mag expected for LSST images, which makes our dataset particularly valuable for studies aimed at forecasting LSST performance. We note that, in spite of the mentioned single-visit depth, we limit our analysis to sources with $r \lesssim 23.5$ mag in order to minimize the inclusion of sources with noise-affected light curves. 
For details on the reduction and combination of exposures, performed using the VST-Tube pipeline~\citep{grado12}, as well as source extraction and sample assembly, we refer the reader to~\cite{decicco15}. The VST-Tube magnitudes are expressed in the AB system.

Our sample contains 20,647 sources detected in at least 50\% of the dataset visits (i.e., having a minimum of 27 points in their light curves), with an average magnitude of $r \leq 23.5$ mag within a $1\arcsec$-radius aperture. For a sub-sample of 2,414 objects there is the availability of labels, and each object is labeled as Star, Galaxy or AGN. Beside that, we have additional sub-classification available for 414 objects: this was obtained from several works from the literature and was already used in \citet{decicco21, decicco22}. Specifically, we have the following labels, even if more than one of them can be associated to the same source: Type 1 (225 objects), Type 2 (122 objects), MIR AGNs (225 objects), variable (259 objects), and X-ray (362 objects). 
These labels reflect the technique used to identify our AGNs LS. Indeed, no AGNs selection technique is complete, and each one presents both strengths and limitations. Type 1 and Type 2 AGNs in this sample were identified via optical spectroscopy, but these sources are a subsample of the X-ray AGNs in this same LS, which come from the \emph{Chandra}-COSMOS Legacy Catalog \citep{marchesi}. Hence, these are X-ray emitting sources with an optical counterpart, and they were classified as either AGNs type on the basis of the presence (Type 1) or absence (Type 2) of broad (i.e., $\geq 2,000$~km s$^{-1}$) emission lines in their spectra. This is quite a traditional criterion and, as such, it is quite strict, while we are now aware that the spectroscopic features of AGNs may change in time \citep[e.g.,][]{macleod16,clagn}; nevertheless, this basic scheme is still widely used to broadly split these sources in two classes. The MIR AGNs sample was obtained via a selection criterion defined in \citet{donley}, where they identify a typical AGN locus in a diagram comparing the two MIR colors $\log(F[8.0]\mu\mbox{m}/F[4.5]\mu\mbox{m})$ and $\log(F[5.8]\mu\mbox{m}/F[3.6]\mu\mbox{m})$. In our work the MIR information was obtained from the  mentioned COSMOS2015 catalog \citep{laigle}, and the variable AGNs sample comes from \citet{decicco21}. See also Section 4 of \citealt{decicco19} for additional details. The most intriguing type of AGNs in the contest of this work are Type 2, since their optical emission is typically harder to detect compared to Type 1 AGNs. There is indeed a general consensus that AGNs possess a disk-like structure, and that their observed properties are, at least in part, the result of orientation effects \citep[e.g.,][]{antonucci,Urry&Padovani}. Emission at different wavelengths arises from physically distinct regions located at varying distances from the central supermassive black hole \citep[e.g.,][and references therein]{netzer}. Type 2 AGNs are viewed approximately edge-on, which implies that the optical/UV emission from the accretion disk -- typically responsible for the optical variability we are interested in -- is at least partially obscured by the infrared-emitting dusty torus located farther out. While other selection methods may be more effective in identifying Type 2 AGNs, optical variability remains a powerful approach due to its ease of application in wide-field surveys.

We refer to the portion of the dataset with labels as the labeled set (LS) and to the remaining part as the unlabeled set (US). Following the approach of~\cite{decicco21}, we made use  of multiple features extracted as described in \cite{sanchez2021alert}, and listed in Table~\ref{tbl:features}.

\begin{table*}[!htbp]
\caption{List of all the features used for the experiments.}
\centering
 \renewcommand\arraystretch{1.2}
 \footnotesize
 \resizebox{\textwidth}{!}{
 \begin{tabular}{l l l}
\hline 
\hline 
Feature & Description & Reference\\	
\hline 
\ \texttt{$A_{SF}$} & rms magnitude difference of the Structure Function (SF), computed over a 1 yr timescale & \citet{Schmidt}\\	
\ \texttt{$\gamma_{SF}$} & Logarithmic gradient of the mean change in magnitude & \citet{Schmidt}\\	
\ \texttt{GP\_DRW\_$\tau$} & Relaxation time $\tau$ (i.e., time necessary for the time series to become uncorrelated), & \citet{graham17}\\	
\ & from a Damped Random Walk (DRW) model for the light curve & \\ 
\ \texttt{GP\_DRW\_$\sigma$} & Variability of the time series at short timescales ($t << \tau$), & \citet{graham17}\\	
\ & from a DRW model for the light curve & \\ 
\ \texttt{ExcessVar} & Measure of the intrinsic variability amplitude & \citet{nandra1997asca}\\	
\ \texttt{P$_{var}$} & Probability that the source is intrinsically variable & \citet{mclaughlin}\\	
\ \texttt{$IAR_\phi$} & Level of autocorrelation using a discrete-time representation of a DRW model & \citet{eyheramendy18}\\	
\hline 
\ \texttt{Amplitude} & Half of the difference between the median of the maximum 5\% and of the minimum & \citet{richards11}\\
\ & 5\% magnitudes & \\
\ \texttt{AndersonDarling} & Test of whether a sample of data comes from a population with a specific distribution & \citet{nun}\\	
\ \texttt{Autocor\_length} & Lag value where the autocorrelation function becomes smaller than $\eta^e$ & \citet{kim11}\\	
\ \texttt{Beyond1Std} & Percentage of points with photometric mag that lie beyond 1$\sigma$ from the mean & \citet{richards11}\\	
\ \texttt{$\eta^e$} & Ratio of the mean of the squares of successive mag differences to the variance & \citet{kim14}\\	
\ & of the light curve & \\ 
\ \texttt{Gskew} & Median-based measure of the skew & -\\	
\ \texttt{LinearTrend} & Slope of a linear fit to the light curve & \citet{richards11}\\	
\ \texttt{MaxSlope} & Maximum absolute magnitude slope between two consecutive observations & \citet{richards11}\\	
\ \texttt{Meanvariance} & Ratio of the standard deviation to the mean magnitude & \citet{nun}\\	
\ \texttt{MedianAbsDev} & Median discrepancy of the data from the median data & \citet{richards11}\\	
\ \texttt{MedianBRP} & Fraction of photometric points within amplitude/10 of the median mag & \citet{richards11}\\	
\ \texttt{PeriodLS} & Period obtained using the \texttt{P4J} Python package (\url{https://github.com/phuijse/P4J}) & \citet{Huijse18}\\	
\ \texttt{PairSlopeTrend} & Fraction of increasing first differences minus the fraction of decreasing first differences & \citet{richards11}\\	
\ & over the last 30 time-sorted mag measures & \\ 
\ \texttt{PercentAmplitude} & Largest percentage difference between either max or min mag and median mag & \citet{richards11}\\	
\ \texttt{Q31} & Difference between the third and the first quartile of the light curve & \citet{kim14}\\	
\ \texttt{Period\_fit} & False-alarm probability of the largest periodogram value obtained with LS & \citet{kim11}\\	
\ \texttt{$\Psi_{CS}$} & Range of a cumulative sum applied to the phase-folded light curve & \citet{kim11}\\	
\ \texttt{$\Psi_\eta$} & $\eta^e$ index calculated from the folded light curve 
 & \citet{kim14}\\	
\ \texttt{R$_{cs}$} & Range of a cumulative sum & \citet{kim11}\\	
\ \texttt{Skew} & Skewness measure & \citet{richards11}\\	
\ \texttt{Std} & Standard deviation of the light curve & \citet{nun}\\	
\ \texttt{StetsonK} & Robust kurtosis measure & \citet{kim11}\\	
\hline 
\ \texttt{class\_star} & \emph{HST} stellarity index & \citet{koekemoer},\\ 
\ & & \citet{scoville}\\ 
\hline 
\ \texttt{u-B} & CFHT $u$ magnitude -- Subaru $B$ magnitude & \citet{laigle}\\ 
\ \texttt{B-r} & Subaru SuprimeCam $B$ mag -- Subaru SuprimeCam $r$+ mag & \citet{laigle}\\ 
\ \texttt{r-i} & Subaru SuprimeCam $r+$ mag -- Subaru SuprimeCam $i+$ mag & \citet{laigle}\\ 
\ \texttt{i-z} & Subaru SuprimeCam $i+$ mag -- Subaru SuprimeCam $z$++ mag & \citet{laigle}\\ 
\ \texttt{z-y} & Subaru SuprimeCam $z$++ mag -- Subaru Hyper-SuprimeCam $y$ mag & \citet{laigle}\\ 
\hline
\ \texttt{Ch21} & \emph{Spitzer} 4.5 $\mu$m (\emph{channel2}) mag -- 3.6 $\mu$m (\emph{channel1}) mag & \citet{laigle}\\
\hline
\end{tabular}\label{tbl:features}
}
\tablefoot{The first two blocks of the table report variability features; \texttt{class\_star} is a  morphology feature, the only one that we used. The bottom part of the table reports the color features used, where \texttt{Ch21} is the only MIR color used, while the others are optical or NIR colors. Table extracted from \cite{decicco21}.}
\end{table*}

\begin{figure}[!htpb]
    \centering

\definecolor{colorAGN}{HTML}{66c2a5}
\definecolor{colorSTAR}{HTML}{fc8d62}
\definecolor{colorGAL}{HTML}{8da0cb}
    \begin{tikzpicture}
        \node[draw, fill=colorAGN, minimum width=0.5cm, minimum height=0.4cm] at (0,0) {};
        \node at (0.8, 0) {AGN};
        \node[draw, fill=colorSTAR, minimum width=0.5cm, minimum height=0.4cm] at (4.4,0) {};
        \node at (5.2, 0) {Star};
        \node[draw, fill=colorGAL, minimum width=0.5cm, minimum height=0.4cm] at (2.2,0) {};
        \node at (3.0, 0) {Gal};
    \end{tikzpicture}
    
    \subfloat[$\eta^{e}$]{\includegraphics[width=0.47\columnwidth, trim=0cm 0cm 0cm 0cm, clip]{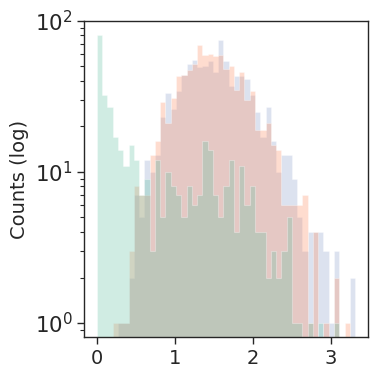}}
    \subfloat[u-B]{\includegraphics[width=0.47\columnwidth, trim=0cm 0cm 0cm 0cm, clip]{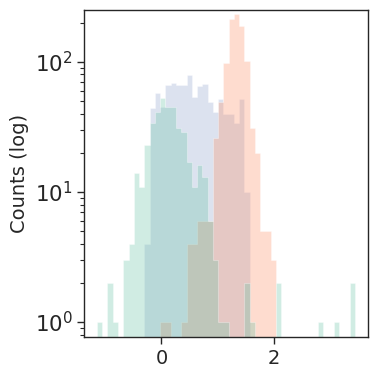}}\\
    \subfloat[StetsonK]{\includegraphics[width=0.47\columnwidth, trim=0cm 0cm 0cm 0cm, clip]{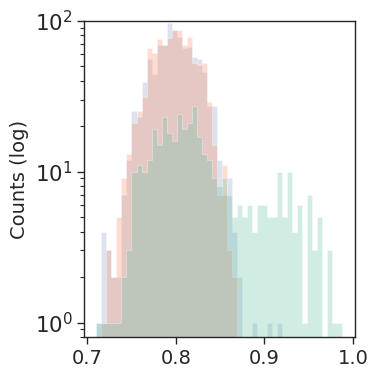}}
    \subfloat[class\_star\_hst]{\includegraphics[width=0.47\columnwidth, trim=0cm 0cm 0cm 0cm, clip]{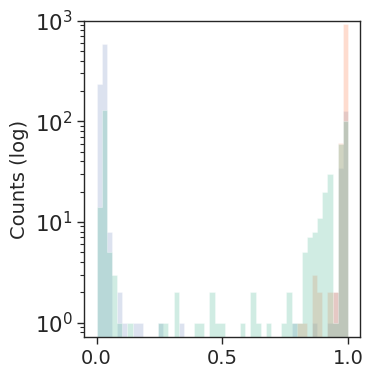}}

  \caption{The distributions of four features in the dataset: $\eta^{e}$, u-B, StetsonK, class\_star\_hst}
 \label{fig:features_distribution}
\end{figure}
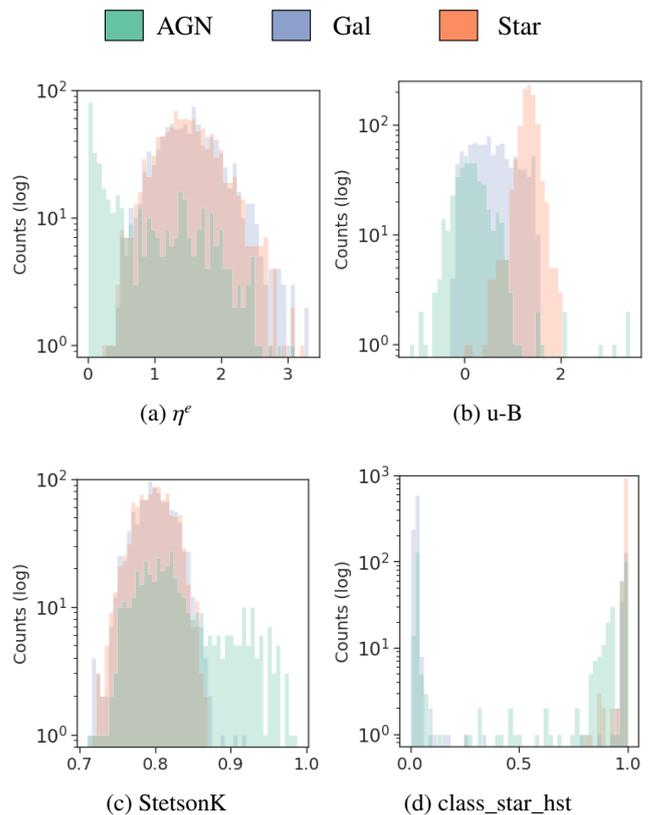

In Figure~\ref{fig:features_distribution} we show the distributions of four selected features: $\eta^{e}$, u-B, StetsonK, class\_star\_hst. These features were chosen to illustrate that, while certain AGNs occupy regions of the parameter space not populated by stars or galaxies, a substantial number of them remain indistinguishable based solely on a limited set of features. This highlights the need to consider the full multi-dimensional feature space, where class separability may be more pronounced.

\section{Method}\label{sec:method}

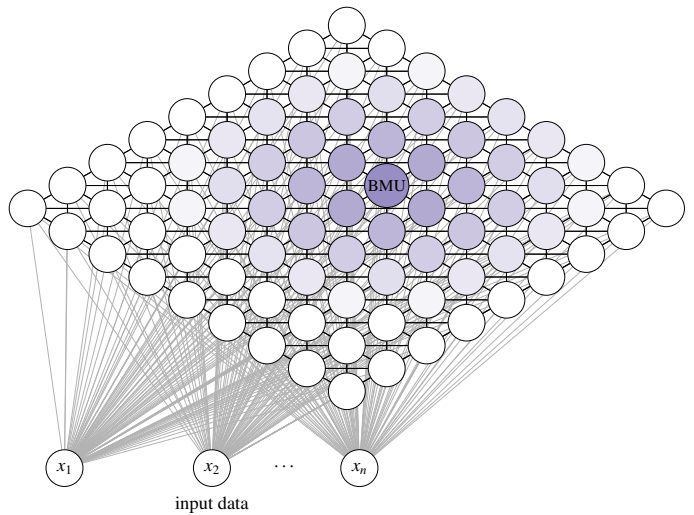
\begin{figure}[!htbp]
\centering

\usetikzlibrary{calc, backgrounds}
\definecolor{colorHit}{HTML}{998ec3}
\resizebox{.49\textwidth}{!}{
\begin{tikzpicture}[
    scale=0.7,
    perspective/.style={x={(0.866cm,-0.5cm)}, y={(0.866cm,0.5cm)}, z={(0cm,1cm)}},
    neuron/.style={circle, draw, minimum size=0.7cm, inner sep=1pt},
    input/.style={circle, draw, minimum size=0.6cm},
]

\begin{scope}[perspective]
   
    \foreach \i in {0,...,8} {
        \foreach \j in {0,...,8} {
            \pgfmathsetmacro{\shade}{100-min(100,max(0,25*sqrt((\i-4)^2+(\j-5)^2)))}
            \ifnum\i=4
                \ifnum\j=5
                    \node[neuron, fill=colorHit] (n\i\j) at (\i*1.25,\j*1.25,0.1) {\small BMU};
                \else
                    \node[neuron, fill=colorHit!\shade] (n\i\j) at (\i*1.25,\j*1.25,0.1) {};
                \fi
            \else
                \node[neuron, fill=colorHit!\shade] (n\i\j) at (\i*1.25,\j*1.25,0.1) {};
            \fi
        }
    }
    
    \foreach \i in {0,...,8} {
        \foreach \j in {0,...,8} {
            \foreach \di/\dj in {-1/-1, -1/0, -1/1, 0/-1, 0/1, 1/-1, 1/0, 1/1} {
                \pgfmathtruncatemacro{\ni}{\i+\di}
                \pgfmathtruncatemacro{\nj}{\j+\dj}
                \ifnum\ni>-1
                    \ifnum\ni<9
                        \ifnum\nj>-1
                            \ifnum\nj<9
    \draw[black, very thin] (n\i\j) -- (n\ni\nj);
                            \fi
                        \fi
                    \fi
                \fi
            }
        }
    }
\end{scope}

\node[input] (x1) at (1, -7) {$x_1$};
\node[input] (x2) at (5, -7) {$x_2$};
\node at (7, -7) {$\cdots$};
\node[input] (xn) at (9, -7) {$x_n$};

\node at (5, -8) {input data};

\begin{pgfonlayer}{background}
    \foreach \x in {x1, x2, xn} {
        \foreach \i in {0,...,8} {
            \foreach \j in {0,...,8} {
                \draw[black!30, very thin] (\x) -- (n\i\j);
            }
        }
    }
\end{pgfonlayer}

\end{tikzpicture}
}
\caption{Schematic representation of the SOM structure, where the input data vector is directly passed to the neurons through the weights, then the adaptation apply to the BMU and its neighbors.}\label{fig:som}
\end{figure}

We used a SOM \citep{kohonen2001basic} to characterize the multidimensional feature space obtained as described in Section~\ref{sec:data}, adopting the python package MiniSOM \citep{vettigliminisom}. A SOM is a helpful tool primarily used for dimensionality reduction and data visualization \citep{kohonen2013essentials}, which belongs to the unsupervised learning domain. It projects the input parameter space onto a lower dimensional grid, a two-dimensional structure in its original form \citep{kohonen2001basic}, although some other topologies exist \citep[e.g.,][]{zin2014cluster}. The grid consists of a $m \times n$ array of nodes, or neurons, as can be seen in Fig.~\ref{fig:som}. Each neuron is represented by a weight vector \textbf{w} that has the same dimension \textit{d} of the input data. Unlike traditional neural networks using backpropagation, the SOM uses competitive learning to represent the dataset. For a given input vector \textbf{x} randomly chosen from the data, the best matching unit (BMU) is found as the neuron that ``wins'' the competition because its weight vector is closest to the input vector, given by:
\begin{equation}
    \text{BMU} = \arg \min_{i} \| \mathbf{x} - \mathbf{w}_i \|~,
\end{equation}
where $\|\cdot \|$ is the Euclidean norm. After finding the BMU, the weight vector of the BMU is updated in order to be dragged closer to the data point, together with the weight vectors of the neurons within a ``neighborhood radius'', which are also updated according to the neighborhood function adopted:
\begin{equation}
    \mathbf{w}_j(t+1) = \mathbf{w}_j(t) + \alpha(t) \cdot h_{j, \text{BMU}}(t) \cdot (\mathbf{x}(t) - \mathbf{w}_j(t))~,
\end{equation}
where $\mathbf{w}_j(t)$ is the weight vector of neuron \textit{j} at time \textit{t}, $\alpha(t)$ is the learning rate at time \textit{t}, $h_{j, \text{BMU}}(t)$ is the neighborhood function, which depends on the distance between neuron \textit{j} and the BMU and decreases over time, $\mathbf{x}(t)$ is the input vector at time \textit{t}. The idea behind this update is that neurons close to the BMU will absorb some of the information provided by the input stimulus, \textbf{x}, through a process of shifting (or migrating) their weights toward the input. This process is repeated for many iterations during which the magnitude of the change depends on how much all neurons are as close as possible to the input data, and decreases also with time. Since the objective of the training process is to position nodes with similar weights close to each other on the map, preserving the topological structure of the input space, for assessing the quality of a SOM it is useful to measure the quantization error and the topographical error. Quantization error is the average distance between the input data points and the corresponding BMU of the map \citep{kohonen2009quantization}:
\begin{equation}
    QE = \frac{1}{N} \sum_{t=1}^{N} \| \mathbf{x}(t) - \mathbf{w}_{\text{BMU}}(t) \|~,
\label{eq:QE}
\end{equation}
where $\mathbf{x}(t)$ represents the input data sample at the training \textit{t}, $\mathbf{w}_{BMU}(t)$ is the weight vector of the BMU associated with the input data $\mathbf{x}(t)$, \textit{N} is the total number of input data, and $\|\cdot \|$ is the Euclidean norm. Lower values of the quantization error indicate a better accuracy of the represented data. Topographic error measures how well the topographic structure of the data is preserved on the map \citep{bauer1999neural, kiviluoto1996topology}. The SOM is expected to maintain the neighborhood relationships of the input data, meaning that if two vectors are close in the input space then they should be mapped into neighboring neurons in the SOM. Mathematically, the topological error is the proportion of input vectors for which the first and the second BMUs are not adjacent in the SOM grid. It is given by the following expression:
\begin{equation}
    TE = \frac{1}{N} \sum_{t=1}^{N} \delta_{\text{top}} \left( \mathbf{x}(t) \right)~,
\label{eq:TE}    
\end{equation}
where \textit{N} is the total number of input vectors, $\delta_{\text{top}} ( \mathbf{x}(t))$ is a step function $\delta_{\text{top}} ( \mathbf{x}(t))=1$ if the closest and the second closest BMU for $x(t)$ are not adjacent, otherwise $\delta_{\text{top}} ( \mathbf{x}(t))=0$. Low topological errors means that the SOM preserves the structure of the input space better \citep{uriarte2005topology}.

The performance of the SOM could be affected by the choice of different hyper-parameters for the training, such as the number of neurons, the number of training iterations (or epochs), the learning rate, and so on. With a smaller number of epochs the SOM may not have enough time to properly adjust the neurons' weights. This could lead to a poor representation of the input data, and both the quantization and topological errors would be high. On the contrary, an excessive number of epochs may lead to overfitting, where the map becomes too finely tuned to the training data, possibly capturing noise or small fluctuations that are not meaningful. Also adopting a large number of nodes the SOM may lead to overfitting, while with a lower number of nodes the SOM may lack sufficient resolution to properly represent the input data and may fail to preserve topological relationships, resulting in high quantization and topological errors. 

As aforementioned, this process is unsupervised. Unlike traditional neural networks, where the weights are optimized in order to match output labels, the SOM learns to differentiate and distinguish features based on similarities, grouping them in a final lower-dimensional space. The presence of labeled data in our dataset is only used for easier interpretation of the SOM results. By identifying the labels of the input data which populate specific neurons of the SOM, one could establish the nature of the clusters, associating them with known categories. For this reason, the SOM can be used also as a tool for visualizing the dataset in a 2D representation, and as a canvas where the features distribution can be mapped on. Furthermore, once the SOM has been trained on the entire dataset, the labeled data can also be used to assign a likely label to each neuron of the map. This allows the neuron labels to be propagated to similar input vectors that have been mapped to the same neurons \citep{Song1996}. In this way, we can benefit of the advantages of both the unsupervised learning (i.e., identifying patterns and structures in the data) and the supervised learning (i.e., label prediction).

\subsection*{Identification of optimal hyper-parameters}
In this section, we discuss the selection of hyper-parameters for the training of our SOM:
\begin{itemize}
    \item the number of epochs;
    \item the neighborhood function;
    \item the size of the map.
\end{itemize}

\begin{figure}[!htpb]
\definecolor{color1}{HTML}{F1A340}
\definecolor{color2}{HTML}{998ec3}
    \begin{tikzpicture}
\begin{axis}[
    xlabel=$x$,
    ylabel=$y$,
    xmin=-4, xmax=4,
    ymin=-0.5, ymax=1.1,
    samples=200,
    smooth,
    domain=-4:4,
]

\addplot[color1, thick] {(1 - x^2) * exp(-x^2/2)};
\addplot[color2, thick] {(1/(0.5*sqrt(2*pi)))*exp((-1/2)*(x/0.5)^2)};

\end{axis}

\begin{scope}[xshift=5cm, yshift=0cm]
    \node[anchor=west] at (-0.8,5) {\textcolor{color1}{\rule{0.5cm}{0.2cm}} Mexican Hat};
    \node[anchor=west] at (-0.8,4.7) {\textcolor{color2}{\rule{0.5cm}{0.2cm}} Gaussian};
\end{scope}

\end{tikzpicture}
\caption{Gaussian and Mexican hat functions.} \label{fig:mexhat}

\end{figure}
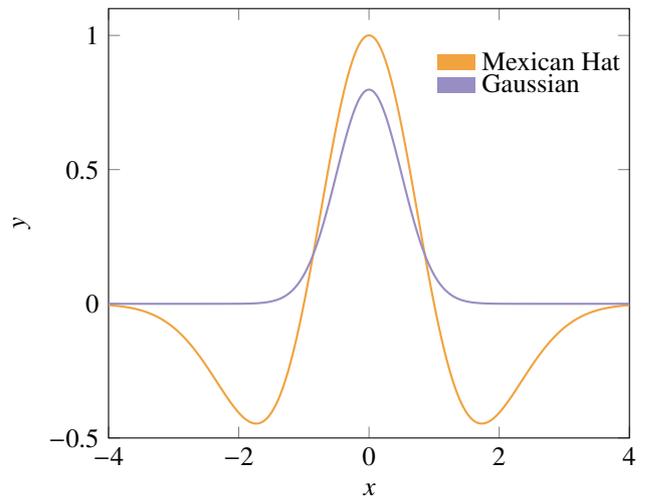

In particular, we performed a grid search to optimize two key hyper-parameters of the SOM: the map size and the number of training epochs. The map size is free to vary in the range 4-60 to explore different levels of resolution in the clustering structure, while the number of epochs ranged from 100 to 2000, in steps of 100, to ensure sufficient convergence without overfitting. To further evaluate the robustness of the training process, the grid search has been conducted using two commonly adopted neighborhood functions: the Gaussian and the Mexican hat (see Fig.~\ref{fig:mexhat}). This systematic exploration allowed us to identify the configuration that provides the most stable and interpretable organization of the input space.

The Gaussian function is a common choice for a neighborhood function, where the influence of the BMU decreases gradually with the distance. Given its shape, this function is particularly beneficial when dealing with data that require a gradual transition between neighboring units, providing a more robust learning process. On the contrary, the Mexican hat: 
\begin{equation}\centering
    h_{j, \text{BMU}}(t) = \left( 1 - \frac{d_{j, \text{BMU}}^{2}}{\sigma(t)^2} \right) \cdot e^{-\frac{d_{j, \text{BMU}}^{2}}{2\sigma(t)^2}}~,
\label{eq:MH}\end{equation} 
where $d_{j, \text{BMU}}$ is the distance between the neuron \textit{j} and the BMU and $\sigma(t)$ is the neighborhood radius that decreases over time, penalizes neighbors that are farther from the BMU: while the neurons close to the BMU are excited, those ones farther away experience negative influence, allowing for sharper boundaries between regions in the map. This function can have some advantages when working with datasets having distinct clusters, as it facilitates a more defined separation between them. However, it can also lead to instability when dealing with a sparse dataset.

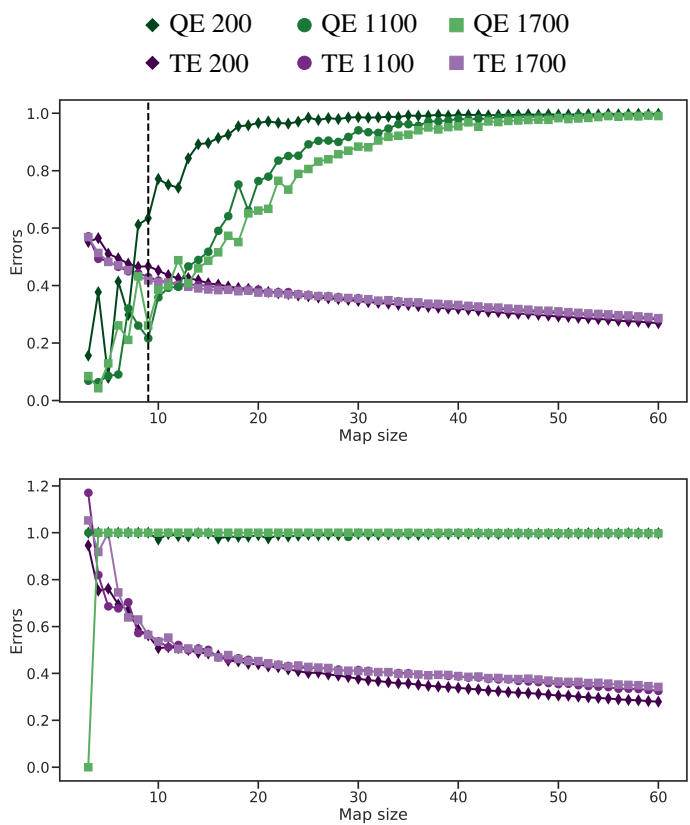
\begin{figure}[!hbpt]
 \centering

 \definecolor{colorTE200}{HTML}{40004b}
 \definecolor{colorTE1100}{HTML}{762a83}
 \definecolor{colorTE1700}{HTML}{9970ab}
 \definecolor{colorQE200}{HTML}{00441b}
 \definecolor{colorQE1100}{HTML}{1b7837}
 \definecolor{colorQE1700}{HTML}{5aae61}

\begin{tikzpicture}
\node[draw=colorQE200, fill=colorQE200, shape=diamond, minimum size=5pt, inner sep=0pt] at (0,0) {};
\node[anchor=west] at (0.1,0) {QE 200};

\node[draw=colorQE1100, fill=colorQE1100, shape=circle, minimum size=5pt, inner sep=0pt] at (2.0,0) {};
\node[anchor=west] at (2.1,0) {QE 1100};

\node[draw=colorQE1700, fill=colorQE1700, shape=rectangle, minimum size=5pt, inner sep=0pt] at (4.0,0) {};
\node[anchor=west] at (4.1,0) {QE 1700};

\node[draw=colorTE200, fill=colorTE200, shape=diamond, minimum size=5pt, inner sep=0pt] at (0,-0.5) {};
\node[anchor=west] at (0.1,-0.5) {TE 200};

\node[draw=colorTE1100, fill=colorTE1100, shape=circle, minimum size=5pt, inner sep=0pt] at (2.0,-0.5) {};
\node[anchor=west] at (2.1,-0.5) {TE 1100};

\node[draw=colorTE1700, fill=colorTE1700, shape=rectangle, minimum size=5pt, inner sep=0pt] at (4.0,-0.5) {};
\node[anchor=west] at (4.1,-0.5) {TE 1700};
\end{tikzpicture}

 \includegraphics[width=1.0\hsize, trim=1.5cm 0cm 0cm 0cm]{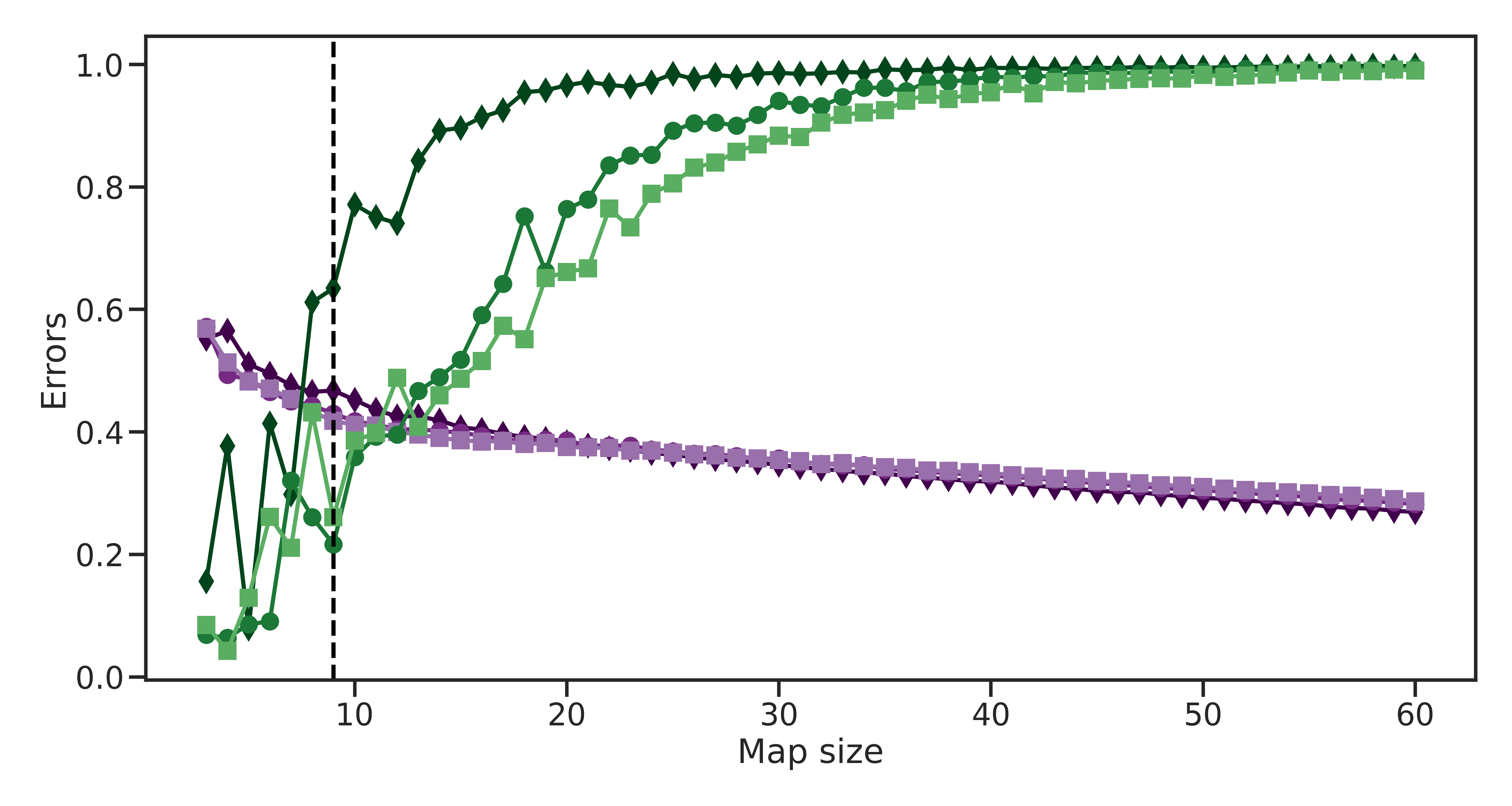}\\
 \includegraphics[width=1.0\hsize, trim=1.5cm 0cm 0cm 0cm]{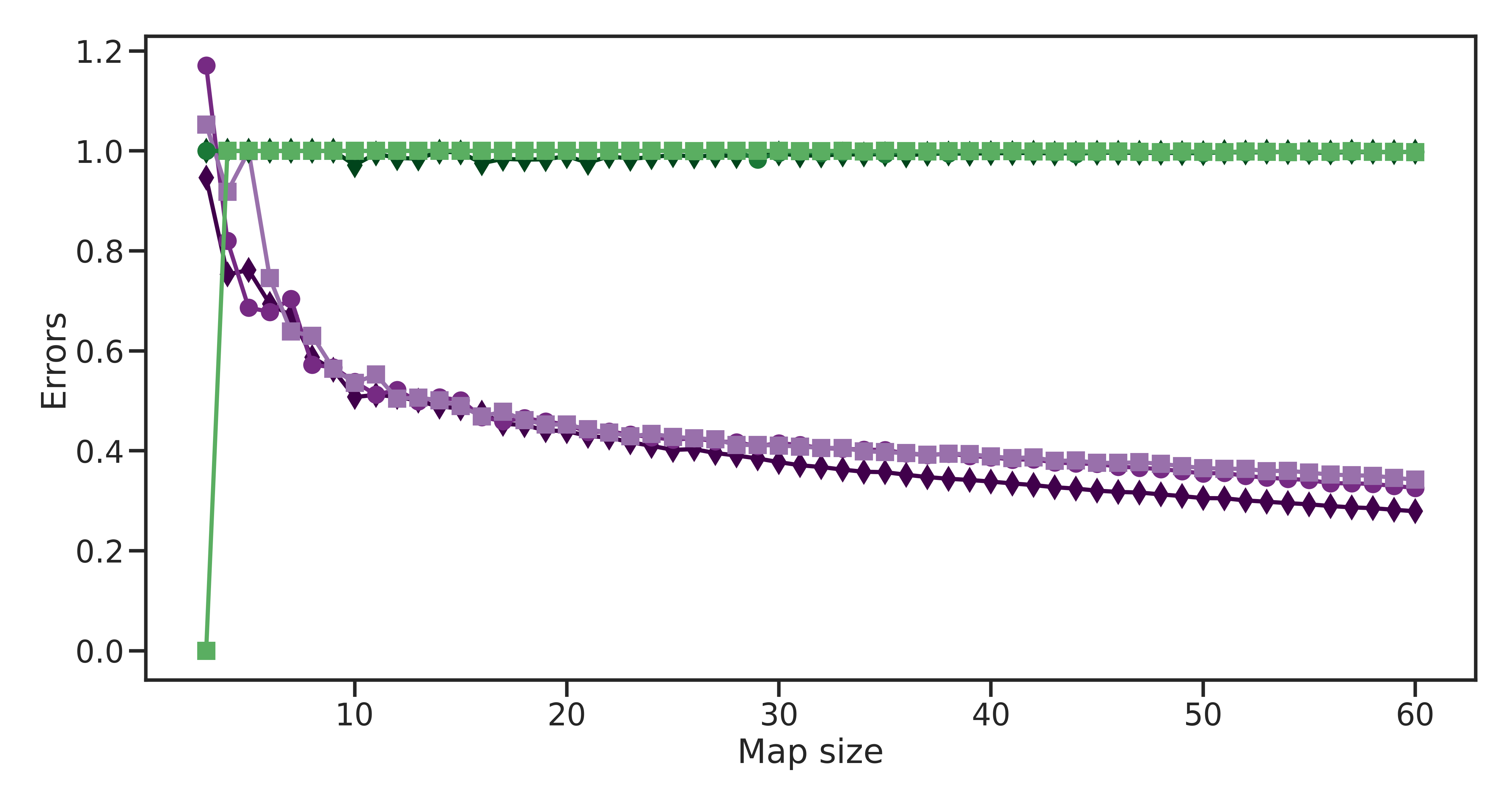}
 
  \caption{Distribution of the quantization error (QE, represented in different shades of green) and topographic error (TE, represented in different shades of purple) as a function of the map size of the SOM, assuming as neighborhood function a Gaussian (top panel) and a Mexican hat (bottom panel). In each plot, the number of training epochs is set to 200 (diamond markers), 1100 (circles) and 1700 (squares), respectively. The dotted line represents the optimal SOM size for representing our dataset with 1100 training epochs.} 
 \label{fig:gridsearch}
\end{figure}

For the purpose of this paper, we do not include all the plots of the results obtained with the grid search. Instead, we show a summary plot reporting the distribution of both quantization and topographic errors as a function of the SOM map size, evaluated for three different numbers of training epochs: 200, 1100, and 1700. These values have been selected to represent different regions of the tested range (100 to 2000 iterations), with 200 and 1700 close to the extremes and 1100 as it has been our final choice. Figure~\ref{fig:gridsearch} displays these results assuming a Gaussian (top panel) and a Mexican hat (bottom panel) as neighborhood function.
It is worth noting that the analysis performed shows that the Mexican hat does not fit well with our starting dataset, and this is particularly evident from the distribution of the topographic error which is almost always constant around one. Therefore, we decided to train a SOM on our dataset adopting a Gaussian function, the number of epochs equal to 1100, and a map size of $9\times9$, since the combination of these parameters seems to preserve the data topology and minimize the distance between data points and their corresponding BMUs.

\section{Experiments/results}\label{sec:results}
In this section we present five experiments on the SOM behavior concerning the usage of different sets of features. The first experiment considers all the variability features with the addition of the optical colors (see Table \ref{tbl:features}). We refer to this experiment as the \textit{Main Experiment}. This initial experiment establishes the reference framework for the construction of the other feature sets, as explained in detail in the following sections.

\subsection{\textit{Main Experiment}}
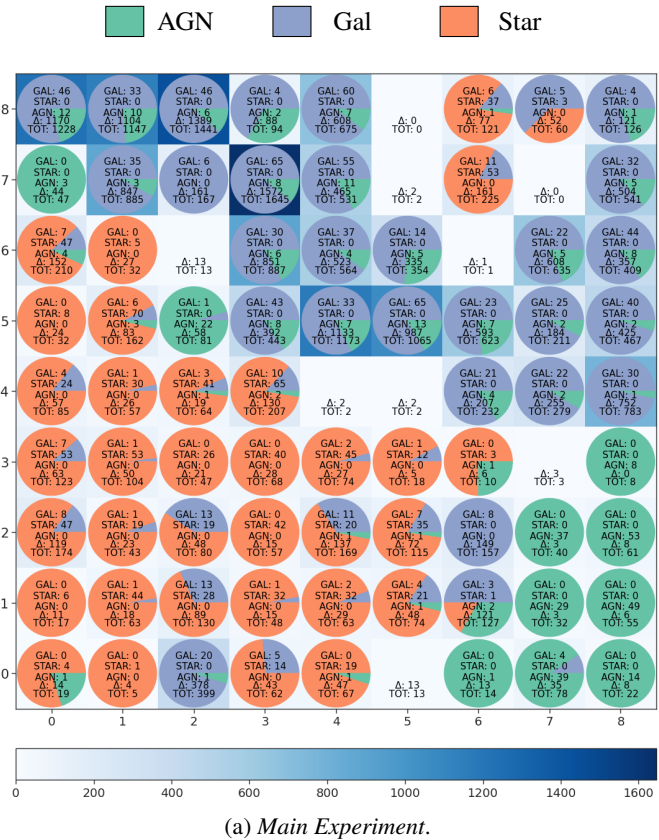
\begin{figure}[!hbpt]
 \centering
 \definecolor{colorAGN}{HTML}{66c2a5}
\definecolor{colorSTAR}{HTML}{fc8d62}
\definecolor{colorGAL}{HTML}{8da0cb}

    \begin{tikzpicture}
        \node[draw, fill=colorAGN, minimum width=0.5cm, minimum height=0.4cm] at (0,0) {};
        \node at (0.8, 0) {AGN};
        \node[draw, fill=colorSTAR, minimum width=0.5cm, minimum height=0.4cm] at (4.4,0) {};
        \node at (5.2, 0) {Star};
        \node[draw, fill=colorGAL, minimum width=0.5cm, minimum height=0.4cm] at (2.2,0) {};
        \node at (3.0, 0) {Gal};
    \end{tikzpicture}
 \subfloat[\textit{Main Experiment}.]{\includegraphics[width=.976\columnwidth,]{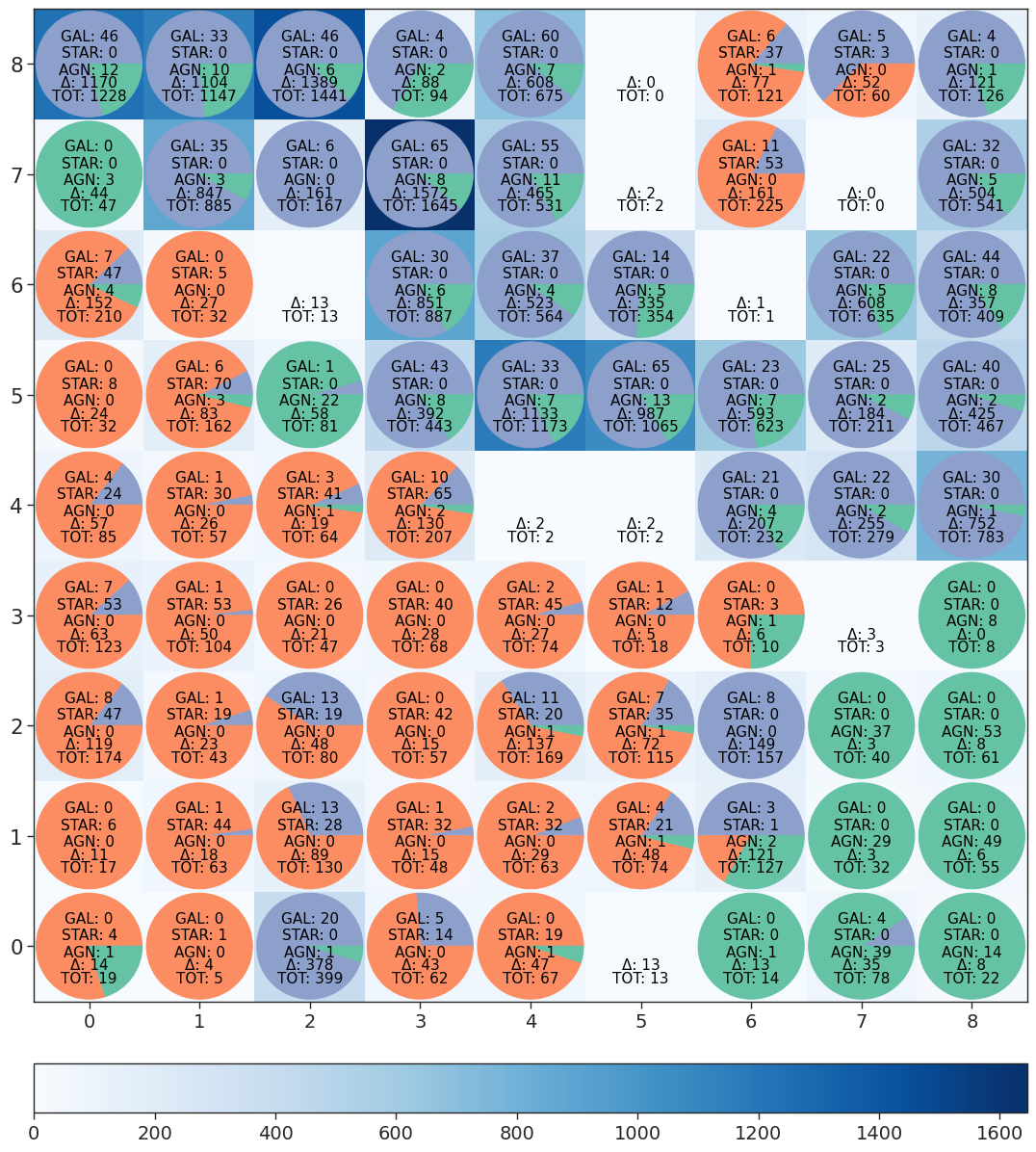}} 
  \caption{Activation map of the \textit{Main Experiment}. A pie chart representing the distribution of labels has been overlaid on each cell of the map. For each cell, it has been indicated: the number of galaxies (GAL), the number of stars (STAR), and the number of AGNs, from the LS; the number of unlabeled objects ($\Delta$); the total number of labeled and unlabeled objects (TOT). The background color intensity reflects the number of objects falling in the cell.} 
 \label{fig:am_labels}
\end{figure}

In this experiment, we ran a SOM on the whole dataset, labeled and unlabeled, using the parameters as explained in Section~\ref{sec:method}, and normalizing it using the MinMaxScaler\footnote{\url{https://scikit-learn.org/1.5/modules/generated/sklearn.preprocessing.MinMaxScaler.html}}. 
The first output of the SOM is the activation map. In this map, each cell corresponds to a specific neuron of the SOM and the value indicates how frequently each neuron is selected as the BMU for the input data. The higher this value, the more times that neuron has been the BMU for some input data, meaning a higher similarity for these data points. On the contrary, lower values may represent outliers or less frequent patterns in the dataset. 
In Figure~\ref{fig:am_labels} we show the activation map obtained for the this set of features\footnote{In this plot and also in all the following ones, it can be noticed that some cells are empty. This means that none of the objects in the dataset choose this cell as BMU.}. We overlaid a pie chart on each cell of the map, representing the distribution of labels for the portion of the dataset where the target is known. This allows one to visualize whether the cells contain objects that share the same label, and understanding the relationship between their features and the labels. Additionally, we specified the total number of the objects (labeled and unlabeled), and the unlabeled objects ($\Delta$) falling into each cell. In this perspective, we can infer that unlabeled objects may receive the same label as those within the same neuron, by assuming that items within the same neuron may share similar characteristics and, therefore, the same label. 

In this Figure, it is possible to observe how most of the known AGNs are distributed in the lower right part of the map, remaining quite uncontaminated by stars and galaxies. Another isolated group of 22 AGNs is positioned in the neuron (2,5)\footnote{We define the notation for neurons on the SOM grid as (column, row).}, contaminated by only one galaxy-type object, while only 3 more AGNs are positioned in the cell (0,7). For the remaining neurons, it is evident that the majority of stars are positioned in the left and central part of the map, while galaxies tend to occupy the opposite space.

\begin{figure*}[!hbpt]

 \centering
 \includegraphics[width=1.0\hsize, trim=0cm 0cm  0cm  0cm, clip]{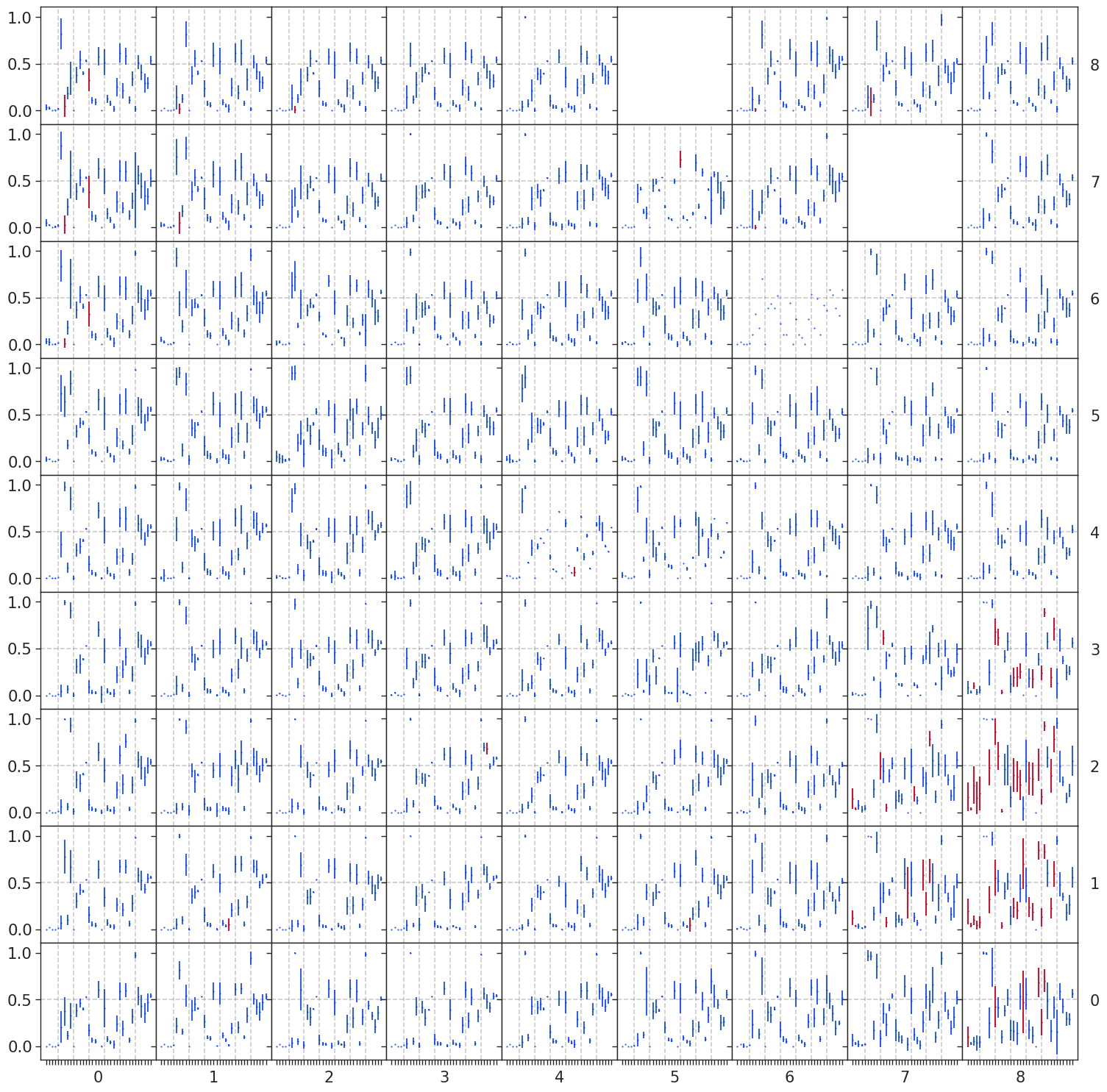}

  \caption{\textit{Main Experiment}. Distribution of the feature means when considering all the features. In each neuron it is represented the local mean of each features, and the error bars correspond to the local standard deviations. The plotted features are in red when their local mean deviates from the corresponding global mean by twice the global standard deviation. The vertical dashed lines are placed after every five features (e.g., at feature 4, 9, 14... 29) to facilitate reading the plot. Order of the features: 0) $A_{SF}$, 1) $\gamma_{SF}$, 2) $GP\_DRW\_\tau$, 3) $GP\_DRW\_\sigma$, 4) ExcessVar, 5) P$_{var}$, 6) $IAR_\phi$, 7) Amplitude, 8) AndersonDarling, 9) Autocor\_length, 10) Beyond1Std, 11) $\eta^{e}$, 12) Gskew, 13) LinearTrend, 14) MaxSlope, 15) Meanvariance, 16) MedianAbsDev, 17) MedianBRP, 18) PeriodLS, 19) PairSlopeTrend, 20) PercentAmplitude, 21) Q31, 22) Period\_fit, 23) $\Psi_{CS}$, 24) $\Psi_{\eta}$, 25) $R_{cs}$, 26) Skew, 27) Std, 28) StetsonK, 29) class\_star\_hst, 30) i-z, 31) r-i, 32) B-r, 33) u-B, 34) z-y.    }
 \label{fig:features_with_colors}

\end{figure*}

It is interesting to investigate how the features are involved within each neuron. For this purpose, we calculated the mean and the standard deviation of each feature within the whole map, which we refer to as the global mean and the global standard deviation. Beside, we calculated the mean and the standard deviation of each feature within the single cell, and which we refer to as the local mean and the local standard deviation. Figure~\ref{fig:features_with_colors} shows the distributions of the feature means (hereinafter, FMD) in each neuron.

In particular, features are plotted in red if the local mean in that cell deviates from the global mean by twice the global standard deviation. A primary observation, based on the comparison between these figures and the corresponding activation maps (in Figure~\ref{fig:am_labels}), is that in all the neurons where AGNs are present, the FMD presents many features in red, meaning that the values of the corresponding objects are highly different and far from the global average. In cells with a majority of AGNs, the photometric colors (which correspond to the last five features) never show a significant difference with respect to the other cells. It is worth noting that the distribution of the mean values of the five colors is quite different for cells with a majority of AGNs, galaxies and stars, respectively. Moreover, there are some features that never become red (thus, they remain within two times the standard deviations from the mean values), as can be seen in the second column of Table~\ref{tbl:neverredfeatures2}, corresponding to the \textit{Main Experiment} row. 
In particular, they are: P$_{var}$, indicating the probability that a source is intrinsically variable, AndersonDarling, indicating whether a sample of data comes from a population with a specific distribution, Skewness (Skew) and its median-based measure (Gskew), the slope of a linear fit to the light curve (LinearTrend), the HST stellarity index (class\_star\_hst), and the four colors i-z, B-r, u-B, z-y. 
The method presented here seems to effectively separate galaxies from stars, as evidenced by Figure~\ref{fig:am_labels}.

\begin{table}[!htbp]
\caption{``Never red features'' and ``Red features''.}
\centering
\setlength{\tabcolsep}{6pt} 
\renewcommand{\arraystretch}{1} 
 \resizebox{\columnwidth}{!}{

\begin{tabular}{lp{3cm}p{3cm}p{3cm}}
\hline\hline
Experiment      & ``Never red'' features & \multicolumn{2}{c}{``Red'' features}   \\
\hline
\multirow{12}{*}{\textit{Main exp.}}          & \multirow{12}{*}{\begin{tabular}[c]{@{}l@{}} P$_{var}$ \\ AndersonDarling \\ Gskew \\ LinearTrend \\ Skew \\ class\_star\_hst \\ i-z \\ B-r \\ u-B \\ z-y \end{tabular}} & $A_{SF}$ \newline $\gamma_{SF}$ \newline $GP\_DRW\_\tau$ \newline $GP\_DRW\_\sigma$ \newline ExcessVar \newline $IAR_\phi$ \newline Amplitude \newline Autocor\_length \newline Beyond1Std \newline $\eta^{e}$ \newline MaxSlope \newline Meanvariance & MedianAbsDev \newline MedianBRP \newline PeriodLS \newline PairSlopeTrend \newline PercentAmplitude \newline Q31 \newline $\Psi_{CS}$ \newline $\Psi_{\eta}$ \newline $R_{cs}$ \newline Std \newline StetsonK \\
\hline
\multirow{8}{*}{\makecell[l]{\textit{Red - corr.} \\\textit{+ top exp.}}}          & \multirow{8}{*}{\begin{tabular}[c]{@{}l@{}} PairSlopeTrend \\ class\_star\_hst \\ P$_{var}$ \\ B-r \\ u-B \end{tabular}} & $A_{SF}$ \newline $\gamma_{SF}$ \newline ExcessVar \newline $IAR_\phi$ \newline Autocor\_length \newline Beyond1Std \newline $\eta^{e}$ \newline MaxSlope \newline MedianAbsDev & MedianBRP \newline PeriodLS \newline Period\_fit \newline $\Psi_{CS}$ \newline $\Psi_{\eta}$ \newline $R_{cs}$ \newline StetsonK \newline r-i \\
\hline
\end{tabular}
}
\tablefoot{Features whose local mean never deviates from the global mean (“never red features”, second column), and those showing deviations (third column), in the respective experiments.}
\label{tbl:neverredfeatures2}
\end{table}

\subsection{Exploring different sets of features}
Starting from the results obtained in the \textit{Main Experiment}, we defined the following feature subsets to enable a more in-depth analysis.
First, we selected as a subset only those features that, in the \textit{Main Experiment}, appeared in red in the FMD. We refer to this subset as the \textit{Red Experiment}.

\begin{figure}[!htbp]
    \centering
    \includegraphics[width=\linewidth]{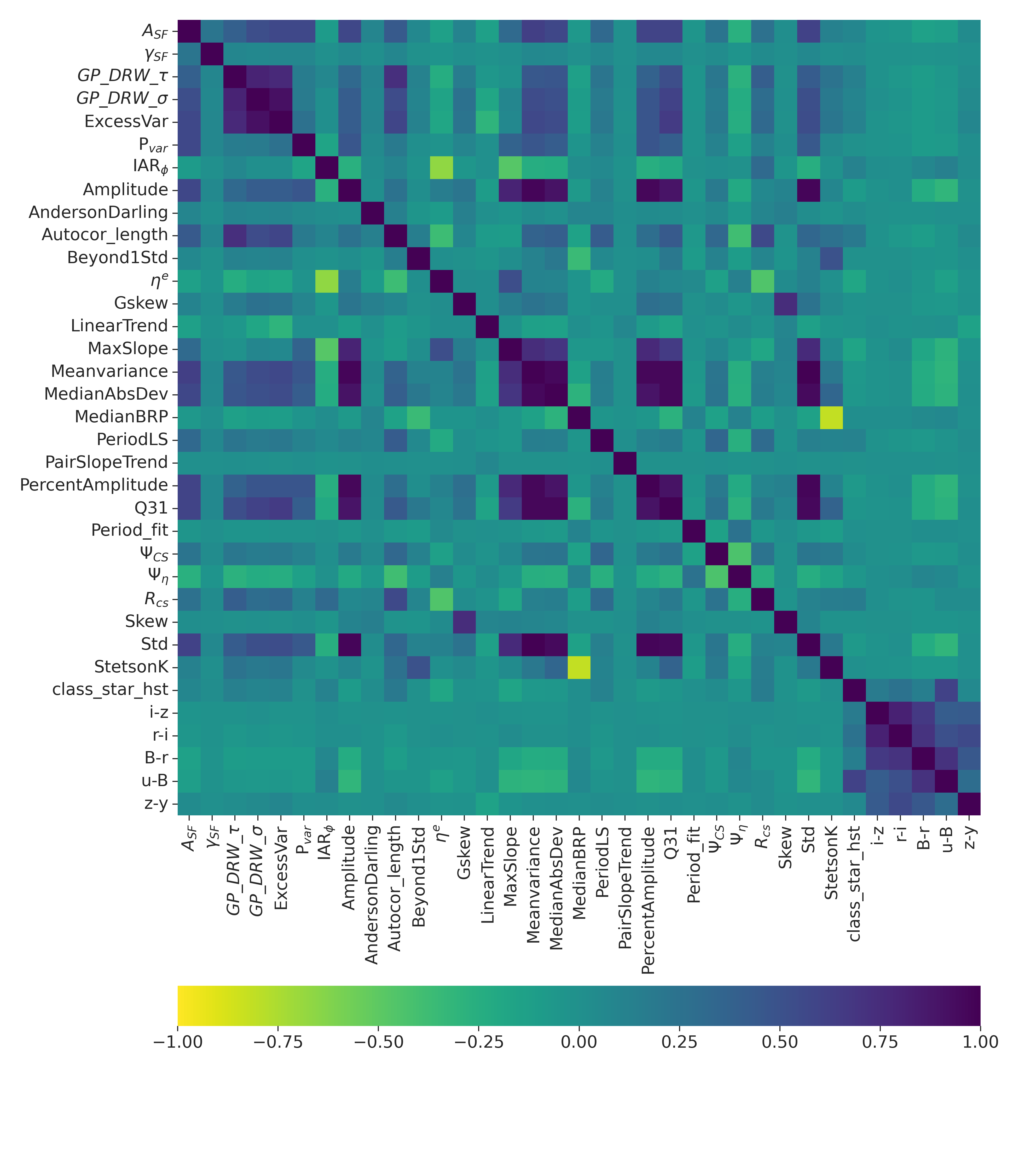}
    \caption{Matrix of correlations among all the features adopted in the\textit{Main Experiment}.}
    \label{fig:correlation}
\end{figure}

Next, we examined the correlations among the features used in the \textit{Main Experiment} (and shown in Fig.~\ref{fig:correlation}) and identified pairs of highly correlated features (absolute Pearson correlation > 90\%). To reduce redundancy, we excluded the following features: PercentAmplitude, Meanvariance, Std, ExcessVar, and retained their correlated counterparts: Q31, Amplitude, MedianAbsDev, GP\_DRW\_sigma. Based on this selection, we defined two additional subsets: \textit{Main - correlated Experiment} and \textit{Red - correlated Experiment}.

Furthermore, we incorporated insights from \cite{decicco19}, where a random forest experiment identified a set of features as particularly relevant for the same dataset. As a fifth feature subset, we considered the union of the \textit{Red - correlated} set and the top ten features from \cite{decicco19}, excluding any features that were among the previously identified highly correlated pairs part of highly correlated pairs. We refer to this combined set as \textit{Red - correlated + top Experiment}.

Finally, we added the feature Ch21 (namely, the color obtained as Spitzer 4.5 $\mu$m (channel2) mag - 3.6 $\mu$m (channel1) mag) as an extra feature to all the previous sets, to analyze its effect on the outcome of the experiments\footnote{We included this color since it appears important in the Feature Importance of \cite{decicco21}.}. The comparative results will provide insights into the importance of colors and Ch21 in the context of AGNs detection.

\begin{figure}[!htbp]
    \centering
    \includegraphics[width=1.0\hsize]{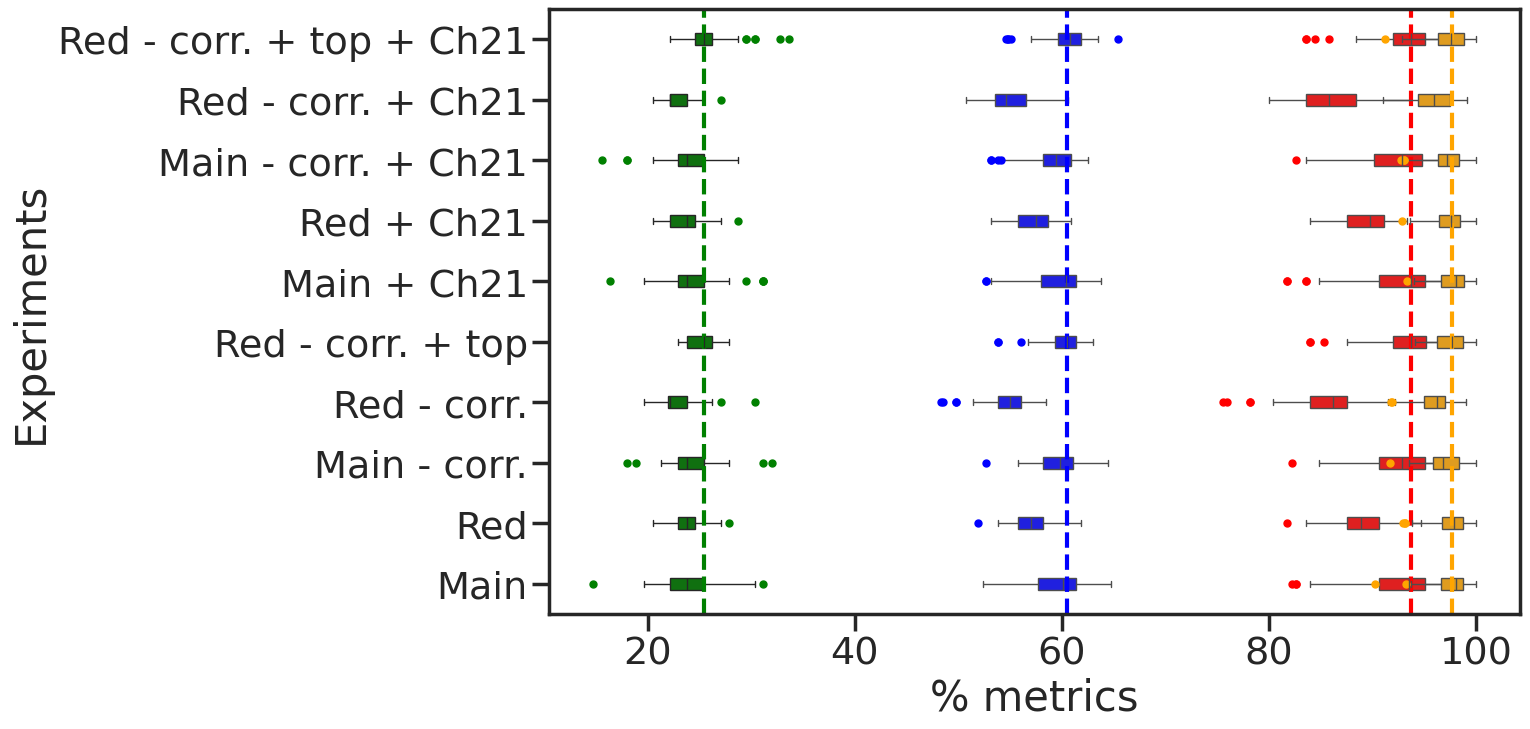}
    \definecolor{complT1}{HTML}{FF0000}
    \definecolor{complT2}{HTML}{008000}
    \definecolor{pureAGN}{HTML}{FFA500}
    \definecolor{complAGN}{HTML}{0000FF}
    \begin{tikzpicture}
        \node[draw, fill=complT1, minimum width=0.15cm, minimum height=0.15cm] at (-0.7,0) {};
        \node at (0.9, 0) {\footnotesize Completeness Type 1};    
        \node[draw, fill=complT2, minimum width=0.15cm, minimum height=0.15cm] at (3.4,0) {};
        \node at (5.0, 0) {\footnotesize Completeness Type 2};
        \node[draw, fill=pureAGN, minimum width=0.15cm, minimum height=0.15cm] at (-0.7,-0.5) {};
        \node at (0.9, -0.5) {\footnotesize Pureness AGNs};        
        \node[draw, fill=complAGN, minimum width=0.15cm, minimum height=0.15cm] at (3.4,-0.5) {};
        \node at (5.0, -0.5) {\footnotesize Completeness AGNs};
    \end{tikzpicture}

    \caption{Distribution of the calculated metrics for each experimental configuration. Each colored box represents one of the four metrics: red for Completeness Type 1, green for Completeness Type 2, orange for Pureness AGNs, and blue for Completeness AGNs. Vertical dashed lines correspond to the median values obtained in the best selected experiment (\textit{Red - corr. + top Experiment}), allowing a direct visual comparison.}
    \label{fig:boxplot} 
\end{figure}

Table \ref{tbl:sets_of_features} resumes the different sets of features adopted for the future experiments.
\begin{table}[!htbp]
\caption{Sets of features adopted in the different experiments.} 
\centering
\setlength{\tabcolsep}{6pt} 
\renewcommand{\arraystretch}{1}
 \resizebox{\columnwidth}{!}{

\begin{tabular}{lp{3cm}p{3cm}}
\hline\hline
Experiment      & \multicolumn{2}{c}{Set of Features}   \\
\hline
\multirow{18}{*}{\shortstack{\textit{Main exp.} \\ \textit{(+ Ch21)}}}          &  $A_{SF}$ \newline $\gamma_{SF}$ \newline  $GP\_DRW\_\tau$ \newline  $GP\_DRW\_\sigma$ \newline  ExcessVar \newline  P$_{var}$ \newline $IAR_\phi$ \newline  Amplitude \newline  AndersonDarling \newline  Autocor\_length \newline Beyond1Std \newline  $\eta^{e}$ \newline  Gskew \newline  LinearTrend \newline  MaxSlope \newline Meanvariance \newline  MedianAbsDev \newline MedianBRP &  PeriodLS \newline PairSlopeTrend \newline  PercentAmplitude \newline  Q31 \newline  Period\_fit \newline $\Psi_{CS}$ \newline  $\Psi_{\eta}$ \newline  $R_{cs}$ \newline  Skew \newline  Std \newline  StetsonK \newline class\_star\_hst \newline  i-z \newline  r-i \newline  B-r \newline  u-B \newline  z-y \newline (Ch21) \\
\hline
\multirow{11}{*}{\shortstack{\textit{Red exp.} \\ \textit{(+ Ch21)}}}          &  $A_{SF}$ \newline $\gamma_{SF}$ \newline $GP\_DRW\_\tau$ \newline $GP\_DRW\_\sigma$ \newline ExcessVar \newline $IAR_\phi$ \newline Amplitude \newline Autocor\_length \newline Beyond1Std \newline $\eta^{e}$ \newline MaxSlope \newline Meanvariance & MedianAbsDev \newline MedianBRP \newline PeriodLS \newline PercentAmplitude \newline Q31 \newline $\Psi_{CS}$ \newline $\Psi_{\eta}$ \newline $R_{cs}$ \newline Std \newline StetsonK \newline (Ch21) \\
\hline
\multirow{14}{*}{\shortstack{\textit{Main - corr. exp.} \\ \textit{(+ Ch21)}}}          &  $A_{SF}$ \newline $\gamma_{SF}$ \newline  ExcessVar \newline  P$_{var}$ \newline $IAR_\phi$ \newline  AndersonDarling \newline  Autocor\_length \newline Beyond1Std \newline  $\eta^{e}$ \newline  Gskew \newline  LinearTrend \newline  MaxSlope \newline  MedianAbsDev \newline MedianBRP \newline PeriodLS & PairSlopeTrend \newline  Period\_fit \newline $\Psi_{CS}$ \newline  $\Psi_{\eta}$ \newline  $R_{cs}$ \newline  Skew \newline  StetsonK \newline class\_star\_hst \newline  i-z \newline  r-i \newline  B-r \newline  u-B \newline  z-y \newline (Ch21) \\
\hline
\multirow{8}{*}{\shortstack{\textit{Red - corr. exp.} \\ \textit{(+ Ch21)}}}          &  $A_{SF}$ \newline $\gamma_{SF}$ \newline ExcessVar \newline $IAR_\phi$ \newline Autocor\_length \newline Beyond1Std \newline $\eta^{e}$ \newline MaxSlope &  MedianAbsDev \newline MedianBRP \newline PeriodLS \newline  $\Psi_{CS}$ \newline $\Psi_{\eta}$ \newline $R_{cs}$ \newline StetsonK \newline (Ch21) \\
\hline
\multirow{11}{*}{\shortstack{\textit{Red - corr. + top exp.} \\ \textit{(+ Ch21)}}}          &  $A_{SF}$ \newline  $\gamma_{SF}$ \newline  ExcessVar \newline  $IAR_\phi$ \newline  Autocor\_length \newline  Beyond1Std \newline  $\eta^{e}$ \newline  MaxSlope \newline  MedianAbsDev \newline  MedianBRP \newline  PeriodLS \newline PairSlopeTrend & Period\_fit \newline  $\Psi_{CS}$ \newline  $\Psi_{\eta}$ \newline  $R_{cs}$ \newline  StetsonK \newline  r-i \newline  u-B \newline  class\_star\_hst \newline P$_{var}$ \newline B-r \newline (Ch21) \\
\hline
\end{tabular}
}
\tablefoot{Ch21 is shown in brackets to indicate its inclusion as an additional feature, in order to evaluate its influence within each set.}
\label{tbl:sets_of_features}
\end{table}
Once the feature subsets have been defined, we ran the SOM on each of the datasets. To ensure the robustness of our results and reduce the impact of random initialization, we generated one hundred different random seeds and used them to initialize the SOM in each run. To objectively evaluate the performance of each experiment, we computed a set of key classification metrics during every run, focusing on four main indicators: Completeness Type 1, Completeness Type 2, Pureness AGNs, and Completeness AGNs.
The boxplot shown in Figure~\ref{fig:boxplot} presents the distribution of metric values across the one hundred SOM seeds used in the analysis, providing an estimate of variability due to initialization. 
From the plot, it is evident that some experimental setups achieve both high median performance and low variability for multiple metrics. 
In contrast, other setups show more pronounced spread or lower overall performance, suggesting less stable or suboptimal behavior under the current feature subsets. 

This comparative visualization clearly highlights the configurations that deliver the most consistent and high-performing results across repeated SOM initializations. Based on a thorough evaluation, we identified \textit{Red - corr. + top} and \textit{Red - corr. + top + Ch21} as the most robust and effective feature sets, and therefore selected them for further analysis.

\subsection{\textit{Red - corr. + top Experiment}}
Once the most suitable feature subsets for our analysis were identified, the next step was to select a representative random seed from those previously generated, in order to obtain results aligned with the average performance. As is shown in the boxplot in Figure~\ref{fig:boxplot}, some seeds lead to either notably higher or lower metric values, potentially biasing the interpretation of the results. To mitigate this effect and ensure a fair representation, we selected $seed = 188$, which produced results closely matching the overall average, to initialize the SOM for subsequent analyses.

\begin{figure}[!htbp]
 \centering
\definecolor{colorAGN}{HTML}{66c2a5}
\definecolor{colorSTAR}{HTML}{fc8d62}
\definecolor{colorGAL}{HTML}{8da0cb}

    \begin{tikzpicture}
        \node[draw, fill=colorAGN, minimum width=0.5cm, minimum height=0.4cm] at (0,0) {};
        \node at (0.8, 0) {AGN};

        \node[draw, fill=colorSTAR, minimum width=0.5cm, minimum height=0.4cm] at (4.4,0) {};
        \node at (5.2, 0) {Star};

        \node[draw, fill=colorGAL, minimum width=0.5cm, minimum height=0.4cm] at (2.2,0) {};
        \node at (3.0, 0) {Gal};
    \end{tikzpicture}

 \includegraphics[width=.976\columnwidth]{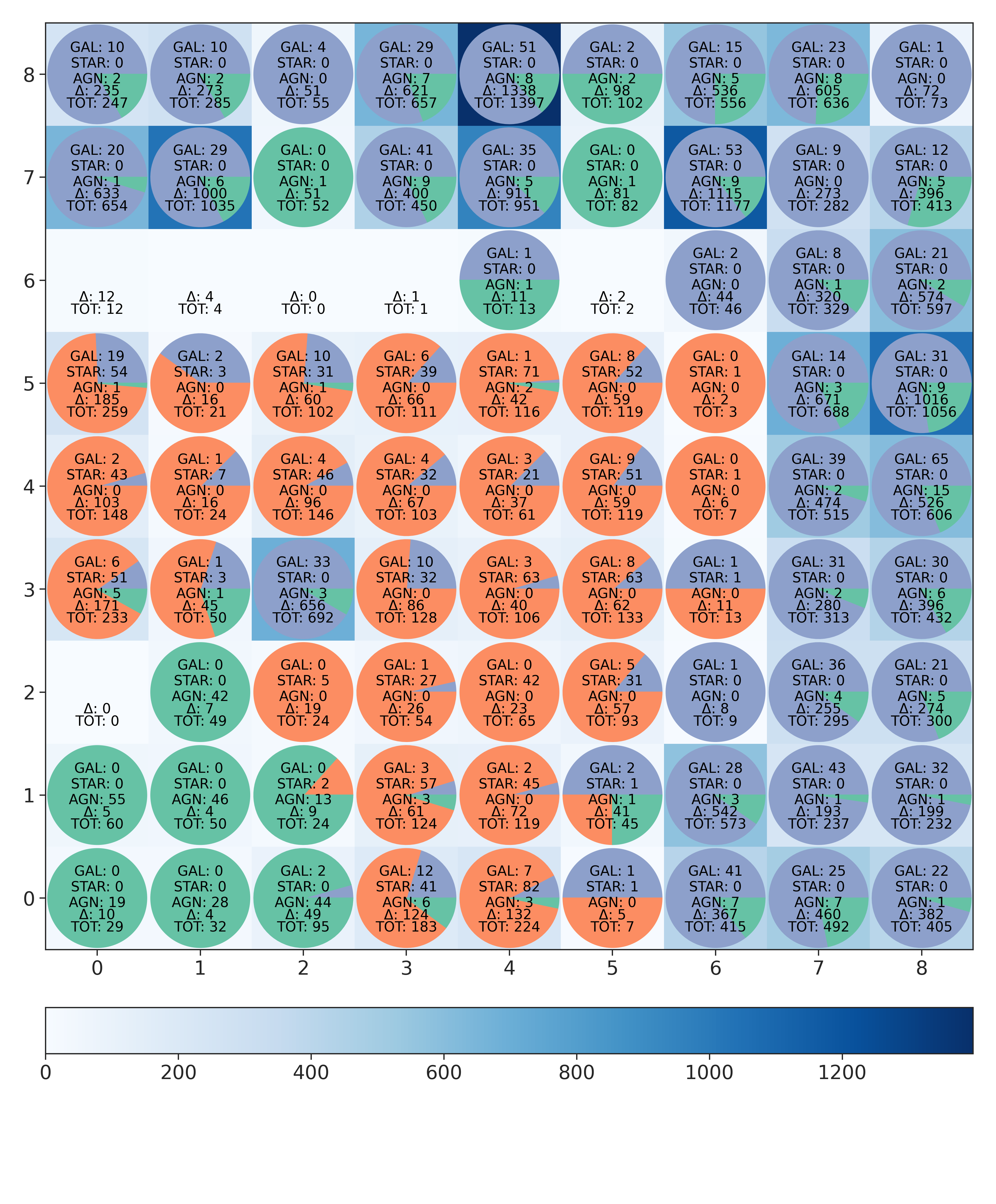} \\
 \includegraphics[width=.976\columnwidth]{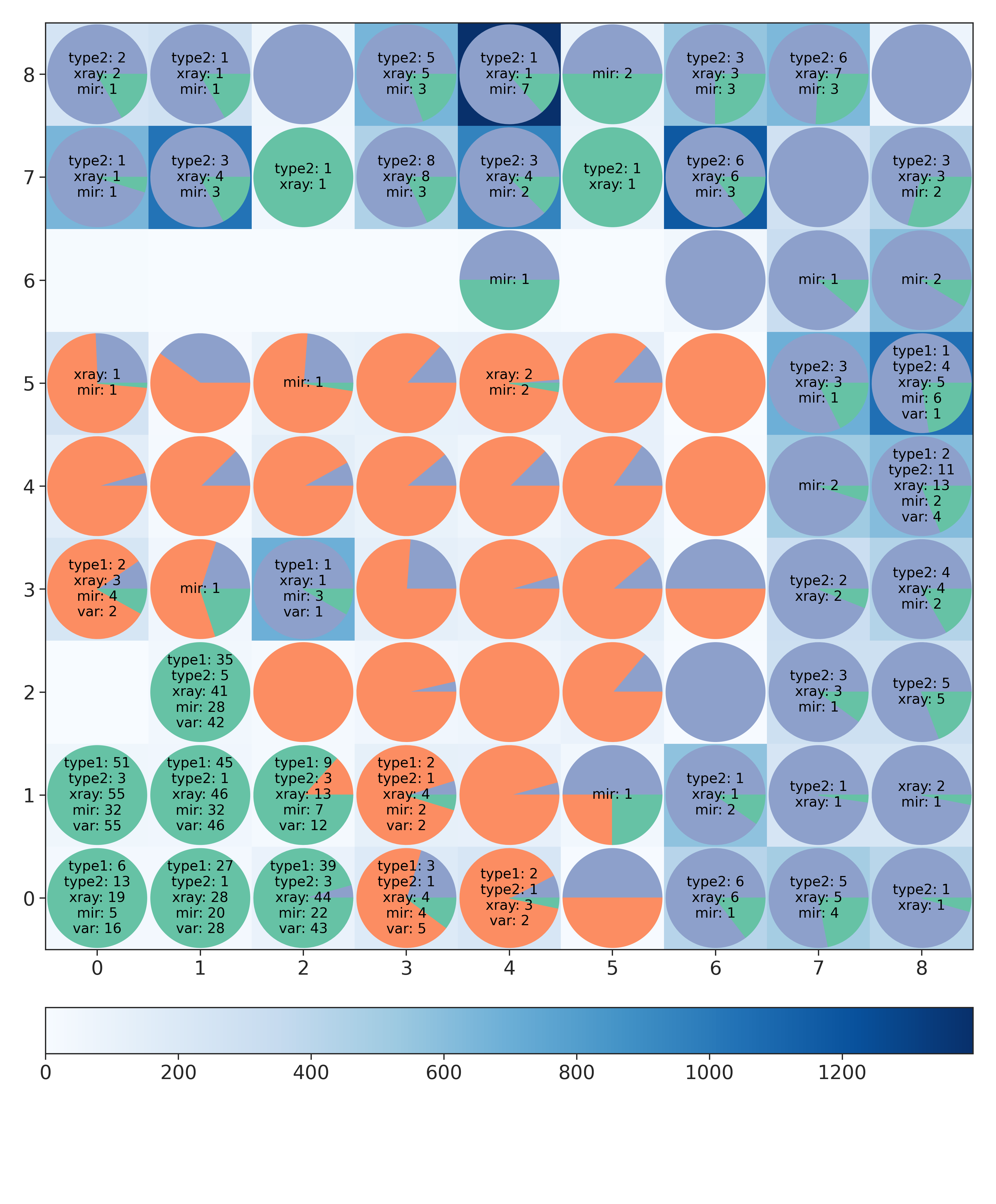}
 \caption{Activation maps of the \textit{Red - corr. + top Experiment}, overlaid by a pie chart representing the distribution of labels (top panel) and the available subclasses for AGNs (bottom panel).}
 \label{fig:red_top_som}
\end{figure}

At this stage, we proceeded to run the SOM on the entire dataset, including both labeled and unlabeled data, using the same parameters adopted in the Main Experiment for consistency. The dataset was normalized using the MinMaxScaler. The top panel of Figure~\ref{fig:red_top_som} shows the activation map obtained in this experiment, where each cell of the map is overlaid by a pie chart, representing the distribution of labels for the portion of the dataset where the target is known. As described in the previous sections, this representation is useful as it reveals distinct clusters that correspond to underlying patterns within the dataset. 

In this panel it can be seen that most of the known AGNs are clearly concentrated in the lower left region of the map, where they form a compact group with minimal contamination from stars and galaxies. This indicates that the SOM has effectively captured the key discriminative features that distinguish AGNs from the other classes, thereby providing a valuable region of interest for identifying potential AGN candidates among the unlabeled data. Beyond this AGN-rich zone, stars are predominantly grouped in well-defined clusters in the central part of the SOM. The distribution of the galaxies, on the other hand, is more scattered  although some degree of local clustering is still observable, indicating some underlying substructure in the input feature space. Among them, four neurons are peculiar for their population: neurons (2,7) and (5,7), have inside only one labeled AGN each, with the rest of the objects being unlabeled. Additionally, neurons (4,6) and (5,8) host only two and four labeled sources, respectively, equally split between AGNs and galaxies, suggesting a more ambiguous region of the map where class boundaries might overlap.

With the bottom panel of Figure~\ref{fig:red_top_som} it is possible to examine our results from another perspective. In this figure, in addition to the pie charts on the activation maps, we have specified the number of AGNs labeled as Type 1, Type 2, X-ray, MIR, and optically variable, respectively.

It is worth noting that, if we examine only cells with a majority of AGNs, Type 1 are quite separated from galaxies, as reported by \cite{decicco21}. In fact, we found that 212 Type 1 objects ($\sim 94\%$) are located in neurons mostly filled by AGNs, while only 31 Type 2 objects ($\sim 25\%$) are found in these cells. It is evident that the percentage of non-AGN objects is quite low ($\sim 2\%$). This indicates that contamination from non-AGN objects is minimal and, while the completeness for Type 2 objects is low, the completeness for Type 1 objects is remarkably high. A summary of these findings is reported in Table~\ref{tab:results} (third column), along with a comparison to the average performance metrics obtained across the 100 random seed runs (second column).

\begin{table}[!htbp]
\setlength{\tabcolsep}{6pt} 
\renewcommand{\arraystretch}{1}
\caption{Summary results of the experiments in terms of completeness of Type 1, Type 2, and purity of AGNs.}
\centering
\resizebox{\columnwidth}{!}{
\begin{tabular}{lcccc}
\hline\hline
    & \multicolumn{2}{c}{\textbf{\textit{Red - corr. + top}}} & \multicolumn{2}{c}{\textbf{\textit{Red - corr. + top + Ch21}}} \\
    & \textbf{mean} & \textbf{s = 188}     & \textbf{mean} & \textbf{s = 2}\\     
    & (\%) & (\%) & (\%) & (\%)\\
\hline
\makecell[c]{\textbf{Completeness}\\\textbf{Type 1}}  & 93.2 $\pm$ 2.6 & 94.2 &  93.2 $\pm$ 2.8 & 93.3 \\
\makecell[c]{\textbf{Completeness}\\\textbf{Type 2}}  & 25.0 $\pm$ 1.4 & 25.4 &  25.5 $\pm$ 2.0 & 25.4 \\
\makecell[c]{\textbf{Purity}\\\textbf{AGNs}}          & 97.5 $\pm$ 1.6 & 98.4 &  97.4 $\pm$ 1.7 & 98.0 \\
\makecell[c]{\textbf{Completeness}\\\textbf{AGNs}}    & 60.3 $\pm$ 1.8 & 60.1 &  60.5 $\pm$ 2.0 & 60.4 \\

\hline
\end{tabular}
}
\label{tab:results}
\end{table}

As was previously discussed, we extended the experiment by including the Spitzer color Ch21 among the features. To ensure a fair comparison, the SOM was trained using the same configuration parameters as in the previous experiment, with the sole exception of the random seed. In this case, the seed was selected to produce results that closely align with the mean performance observed across the one hundred random initializations. For this purpose, we adopted $seed = 2$. In particular, 210 Type 1 objects ($\sim 93\%$) are located in neurons predominantly populated by AGNs, and only 31 Type 2 objects ($\sim 25\%$) are found in these cells. As in the previous experiment, the fraction of non-AGN objects remains low ($\sim 2\%$), suggesting a low contamination from stars and normal galaxies. These results are summarized in the last two columns of Table~\ref{tab:results}, where it is evident that the inclusion of Ch21 in the feature set does not lead to any significant improvement in the results. On the contrary, the metrics remain essentially unchanged, suggesting that the SOM does not exploit this additional feature to discriminate the different objects.

\begin{figure*}[!hbpt]
 \centering
 \includegraphics[width=1.0\hsize, trim=0cm 0cm  0cm  0cm, clip]{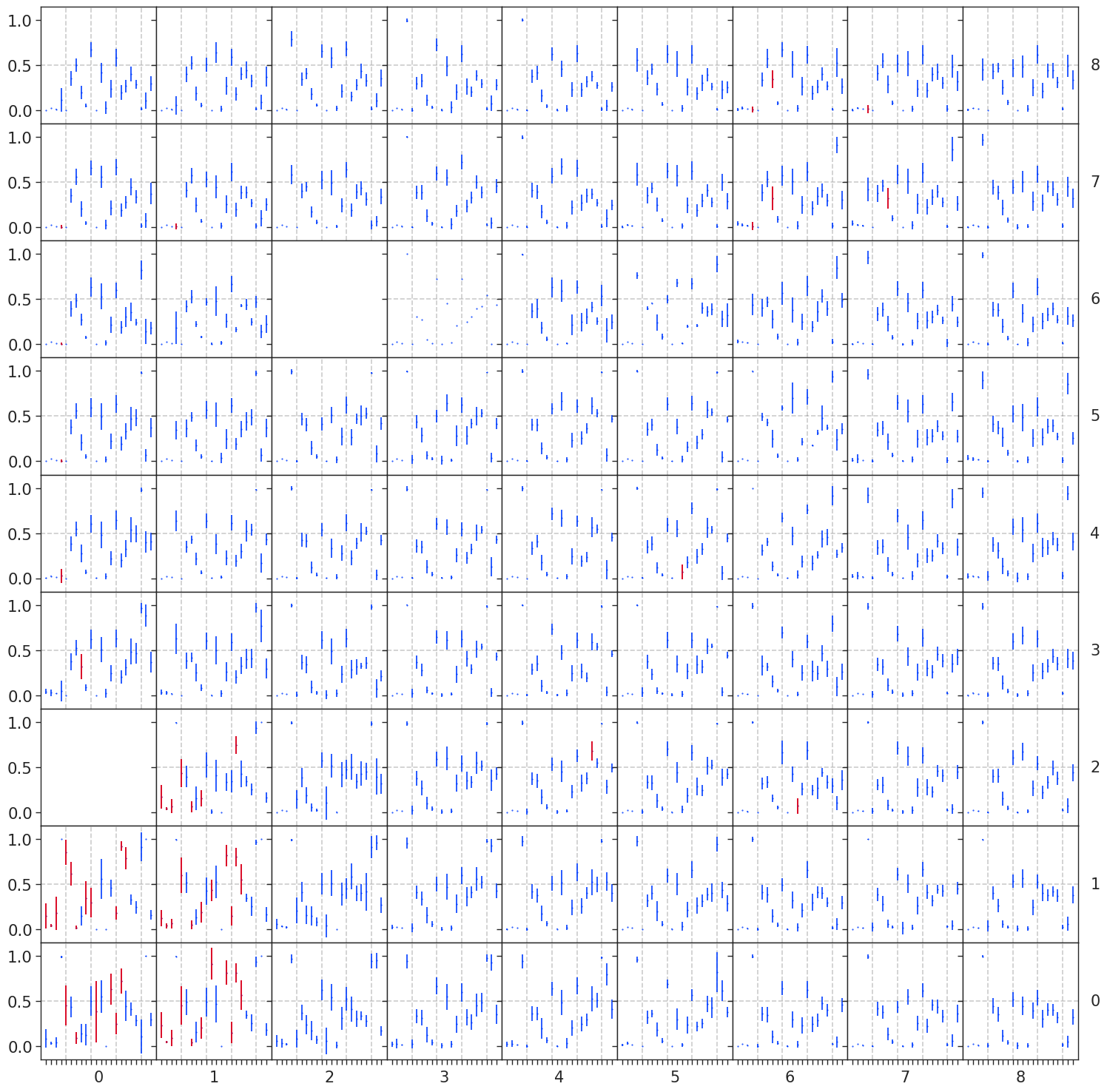}
 
  \caption{\textit{Red - corr. + top Experiment}. Distribution of the feature means (FMD). In each neuron it is represented the local mean of each features, and the error bars correspond to the local standard deviations. The plotted features are in red when their local mean deviates from the corresponding global mean by twice the global standard deviation. The vertical dashed lines are placed after every five features (e.g., at feature 4, 9, 14, 19) to facilitate reading the plot. Order of the features: 0) $A_{SF}$, 1) $\gamma_{SF}$, 2) ExcessVar, 3) $IAR_\phi$, 4) Autocor\_length, 5) Beyond1Std, 6) $\eta^{e}$, 7) MaxSlope, 8) MedianAbsDev, 9) MedianBRP, 10) PeriodLS, 11) PairSlopeTrend, 12) Period\_fit, 13) $\Psi_{CS}$, 14) $\Psi_{\eta}$, 15) $R_{cs}$, 16) StetsonK, 17) r-i, 18) u-B, 19) class\_star\_hst, 20) P$_{var}$, 21) B-r. }  
 \label{fig:features_red_top}
\end{figure*}

In this case we also built the FMD in order to verify the presence or absence of red features, i.e., those features whose values of the corresponding objects are strongly different and far from the global mean. As can be seen from the Figure~\ref{fig:features_red_top}, the red features are present in the cells with a majority of AGNs. The reader should note that this experiment had already started with the red features obtained from the \textit{Main Experiment} (i.e., the one that included all the features at our disposal). Furthermore, only a group of this subset continues to become red, as can be seen in the Table~\ref{tbl:neverredfeatures2} in correspondence of the \textit{Red - corr. + top Experiment}.

After excluding the highly correlated features from the second feature set, and considering the addition of the most relevant features identified by \cite{decicco21}, several observations can be made. First, when comparing the features that never turn red in both experiments, it becomes evident that the color indices B–r and u–B consistently fall below the threshold used for highlighting features in red. The same can also be said for P$_{var}$ and class\_star\_hst, which similarly do not appear to reach the activation levels required for red marking in this context.
Moreover, when using the reduced feature set, we observe that PairSlopeTrend also becomes a feature that is never marked as red. Lastly, within the group of red features in common across both the experiments, Period\_fit and the color r–i emerge as shared key indicators, further reinforcing their relevance in the separation of the different object types.

\begin{figure}[!hbpt]
 \centering
 \definecolor{colorAGN}{HTML}{66c2a5}
 \definecolor{colorSTAR}{HTML}{fc8d62}
 \definecolor{colorGAL}{HTML}{8da0cb}
 \definecolor{colorUS}{HTML}{808080}

    \begin{tikzpicture}
        \node[draw, fill=colorAGN, minimum width=0.5cm, minimum height=0.4cm] at (0,0) {};
        \node at (0.8, 0) {AGN};
        \node[draw, fill=colorSTAR, minimum width=0.5cm, minimum height=0.4cm] at (2.2,0) {};
        \node at (3.0, 0) {STAR};
        \node[draw, fill=colorGAL, minimum width=0.5cm, minimum height=0.4cm] at (4.4,0) {};
        \node at (5.2, 0) {GAL};
        \node[draw, fill=colorUS, minimum width=0.5cm, minimum height=0.4cm] at (6.6,0) {};
        \node at (7.4, 0) {US};
    \end{tikzpicture}
    
\subfloat[Cell (5,7)]{\includegraphics[width=0.361\columnwidth]{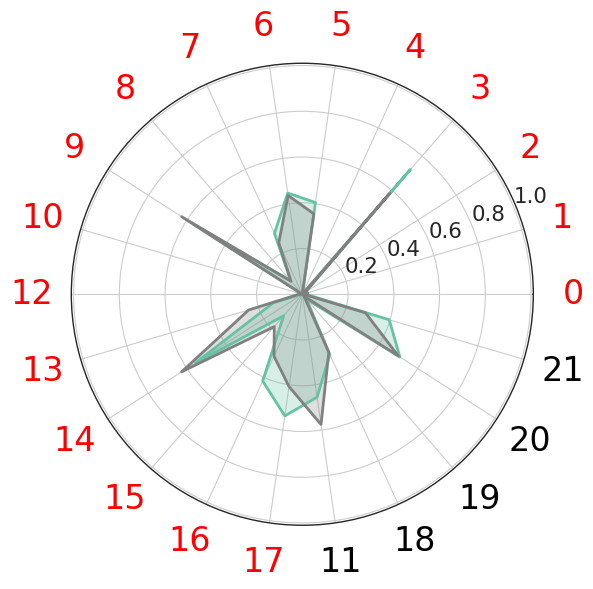}} \hspace{1cm}
\subfloat[Cell (2,7)]{\includegraphics[width=0.361\columnwidth]{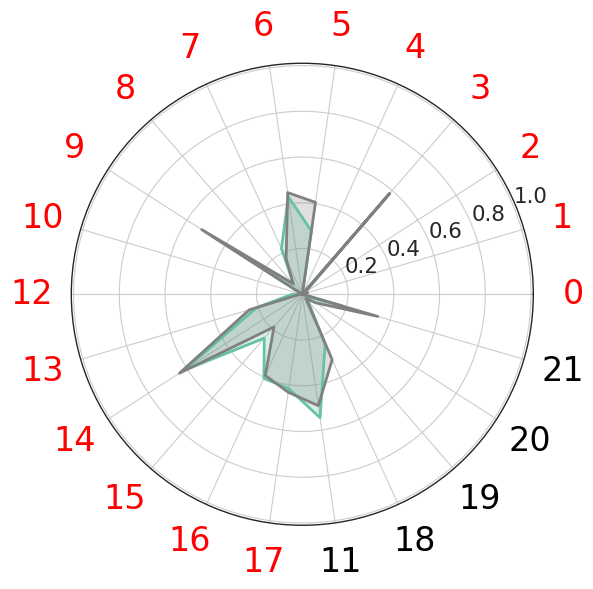}}\\
\begin{tikzpicture}
  \draw[black, thick, dashed] (0,0) -- (8.4,0);
\end{tikzpicture}
\subfloat[Cell (2,1)]{\includegraphics[width=0.361\columnwidth]{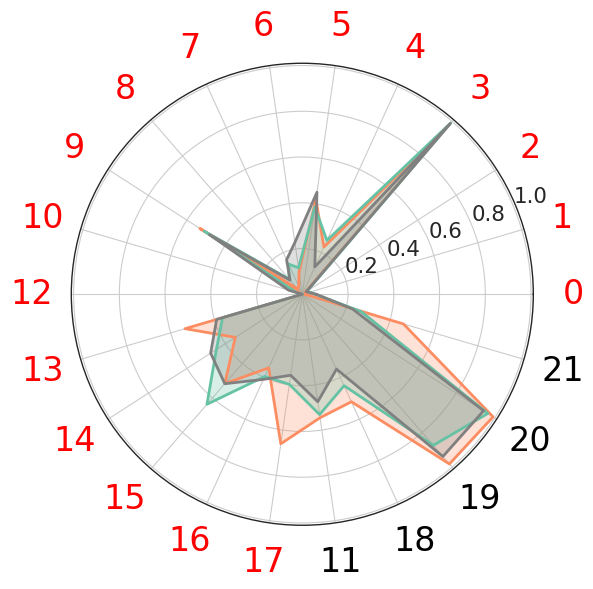}} \\ 
\subfloat[Cell (2,0)]{\includegraphics[width=0.361\columnwidth]{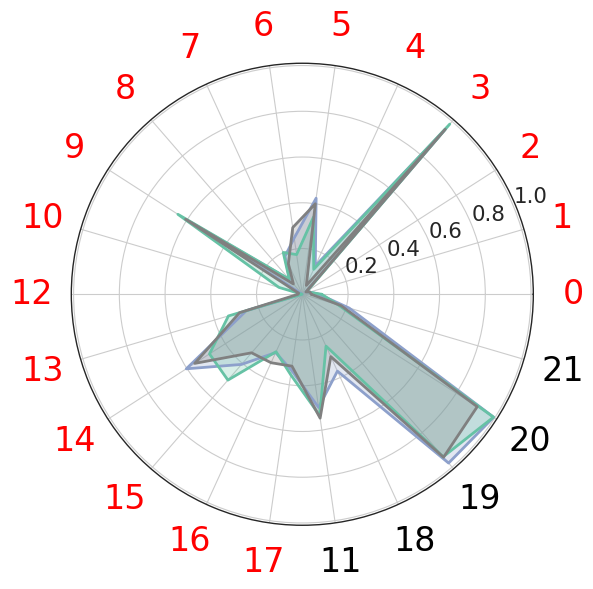}} \hspace{1cm}
\subfloat[Cell (1,2)]{\includegraphics[width=0.361\columnwidth]{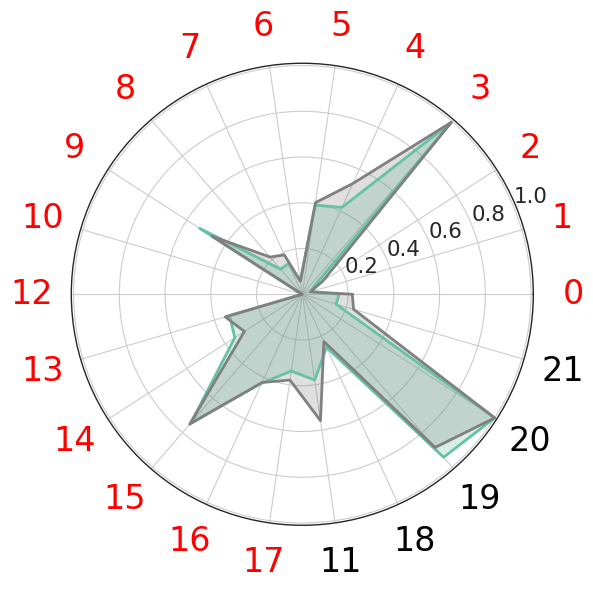}}\\
\subfloat[Cell (0,1)]{\includegraphics[width=0.361\columnwidth]{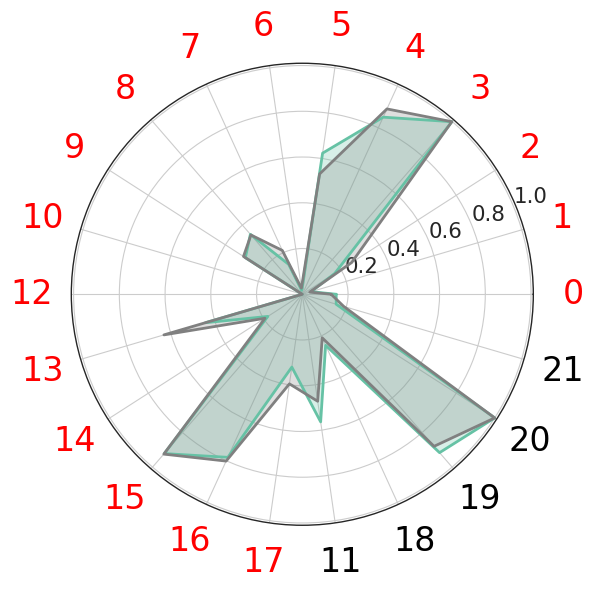}} \hspace{1cm}
\subfloat[Cell (1,1)]{\includegraphics[width=0.361\columnwidth]{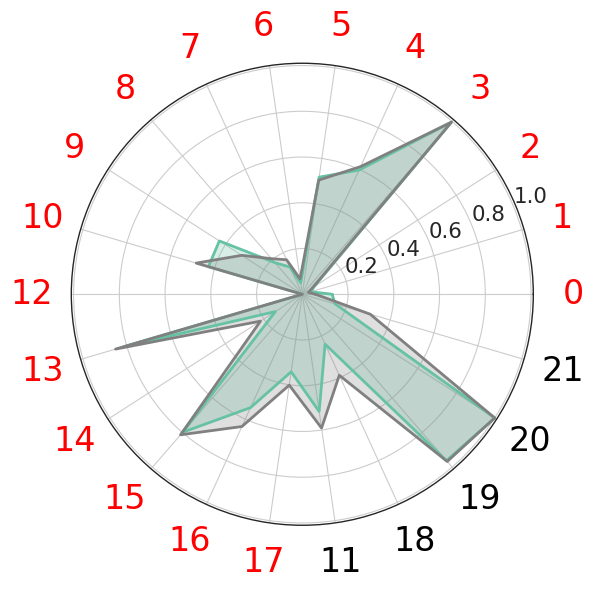}}\\
\subfloat[Cell (0,0)]{\includegraphics[width=0.361\columnwidth]{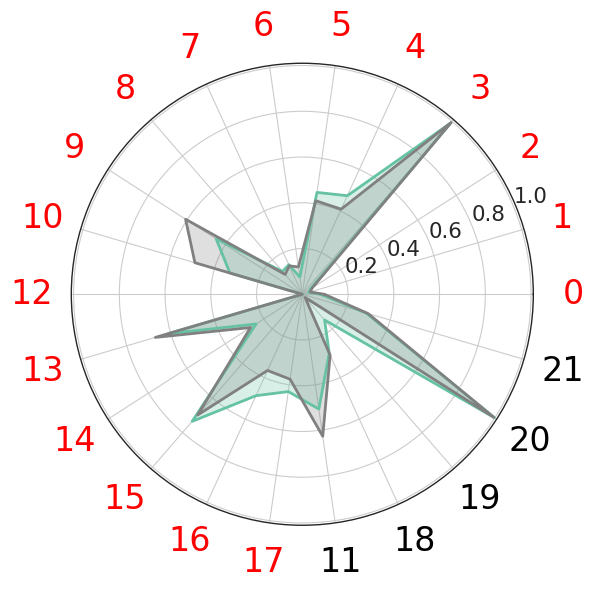}} \hspace{1cm}
\subfloat[Cell (1,0)]{\includegraphics[width=0.361\columnwidth]{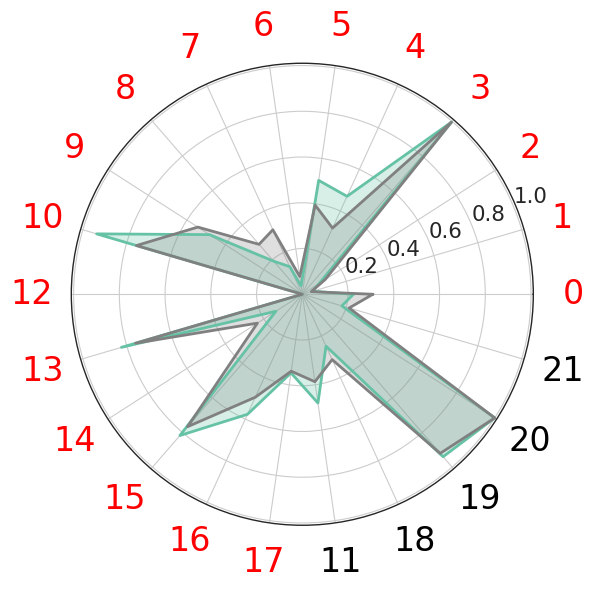}} 
 
  \caption{Distribution of the mean values per neuron of both red and never red features. Results are reported for neurons with an AGNs majority. The top cells (2,7) and (5,7) are separated by the dotted line as they each contain only one labeled AGN. The reader should note that the feature order is chosen to separate red from never-red features. Indexes of features: 0) $A_{SF}$, 1) $\gamma_{SF}$, 2) ExcessVar, 3) $IAR_\phi$, 4) Autocor\_length, 5) Beyond1Std, 6) $\eta^{e}$, 7) MaxSlope, 8) MedianAbsDev, 9) MedianBRP, 10) PeriodLS, 11) PairSlopeTrend, 12) Period\_fit, 13) $\Psi_{CS}$, 14) $\Psi_{\eta}$, 15) $R_{cs}$, 16) StetsonK, 17) r-i, 18) u-B, 19) class\_star\_hst, 20) P$_{var}$, 21) B-r.
  }
 \label{fig:radarplot}
\end{figure}
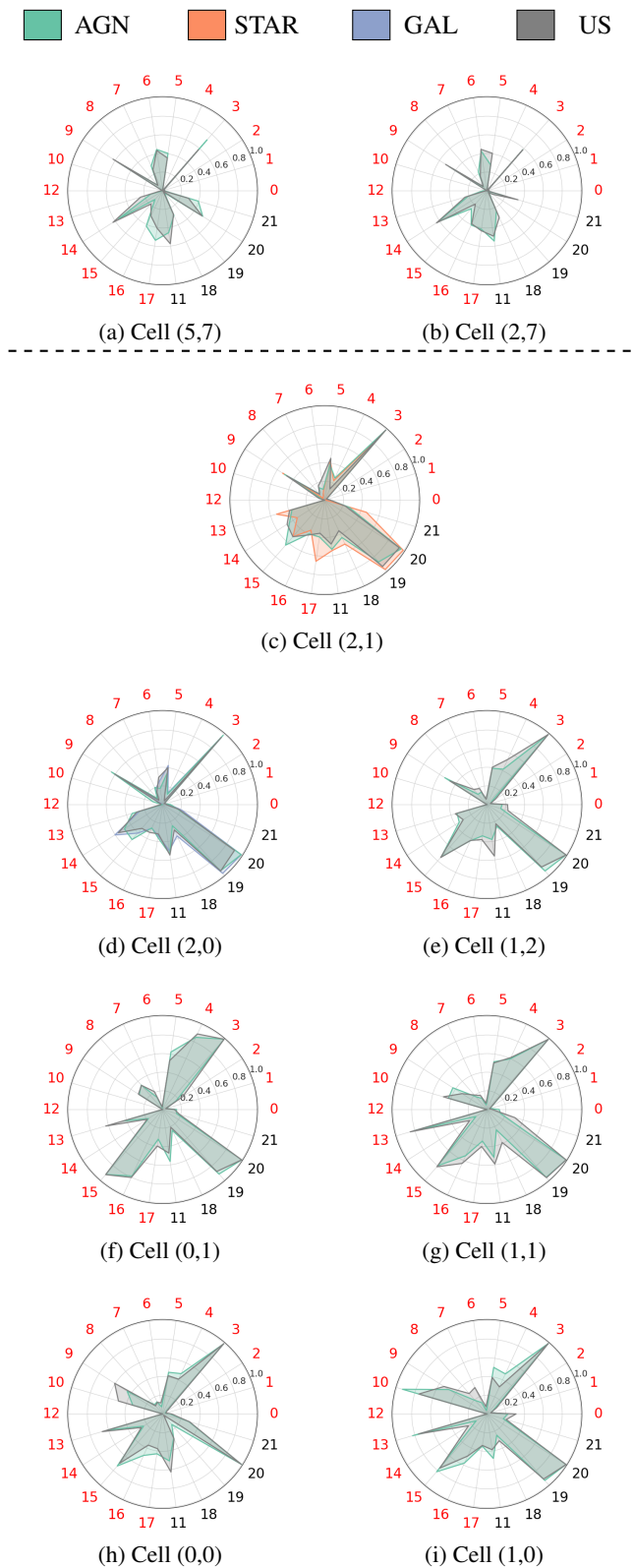

For a better evaluation and analysis of these results, in Figure~\ref{fig:radarplot} we show the distribution of the mean values of both red and never red features, for cells mostly populated by AGNs. Features of neighboring cells have similar behaviors as expected from a SOM. However, neuron (0,0), which has a large fraction of Type 2, shows significant differences in some features. In particular, the objects populating this neuron show a significantly smaller value of class\_star\_hst (feature 19 in the radar plot) than the others. From an observational point of view, Type 2 sources are generally more extended, which is likely a selection effect, since the nucleus is often obscured by the surrounding dust near the accretion disk, limiting our ability to observe it clearly at high redshift. On the contrary, this feature exhibits high values for both AGNs- and stars- dominated cells. Moreover, the P$_{var}$ (feature 20) shows exceptionally high values in almost all cells with an AGNs majority, reflecting their inherent and significant luminosity variability driven by accretion processes around their supermassive black holes. Conversely, we observed very low values of P$_{var}$ for both stars and galaxies, as these objects are typically considered photometrically stable. Exceptions to these typical P$_{var}$ values observed in AGNs-dominated cells are found in neuron (2,7), and to a lesser extent in neuron (5,7). It is worth noting that neurons (2,7) and (5,7) each contain only one labeled AGN, implying limited statistical significance in evaluating their labeled composition. Nevertheless, from the same cells it can be noted how the behavior of the unlabeled objects is completely similar to the single labeled AGN. This highlights a key strength of the SOM as an unsupervised method: it allows one to capture and explore such patterns and similarities regardless of the availability of labeled data. A supervised approach, in contrast, would likely have required a larger number of labeled examples to recognize or validate these associations. Both cells exhibit feature patterns that are atypical from the rest of AGNs-dominated neurons, such as the missing structure in correspondence of $\psi_{CS}$ (feature 13), $\psi_{\eta}$ (feature 14), and $R_{cs}$ (feature 15). Most of the AGNs-dominated neurons shows, in fact, low values of $\psi_{\eta}$, a quite different behavior from the other cells in which stars and galaxies are dominant, suggesting that AGNs dominated sources typically exhibit less short-term variability ($\psi_{\eta}$ low) but more coherent and wide-ranging long-term changes ($\psi_{CS}$ and $R_{cs}$ high) compared to others. Finally, neuron (5,7) shows higher values for the color r-i (feature 17), also observed in other star-dominated neurons (see, for example, the star component represented in orange in cell (2,1). This suggests that while AGNs typically exhibit bluer colors, these specific neurons are sensitive to objects where the contribution from redder and cooler populations is more prominent, further supporting the idea that these cells host peculiar sources isolated by the SOM.

The method presented here seems to effectively separate galaxies from stars, as evidenced by Figure~\ref{fig:red_top_som}. Remarkably, this separation persists even without incorporating the Spitzer color Ch21 information in the training process, suggesting that we can perform the classification based on variability and morphological features alone.

\subsubsection{AGNs ``stability'' within the SOM}
To strengthen the robustness of the results obtained in the \textit{Red - corr. + top Experiment}, we conducted an analysis to determine which kind of objects consistently populate the cells dominated by AGNs. This assessment serves a dual purpose: first, to evaluate the stability and consistency of the experiment; second, to assess whether the presence of objects in the AGNs-majority cells is due to a systematic pattern or merely to random association.

\begin{figure}[!htbp]
    \centering
    \subfloat[N. exp $\ge$ 90]{\includegraphics[width=.5\columnwidth]{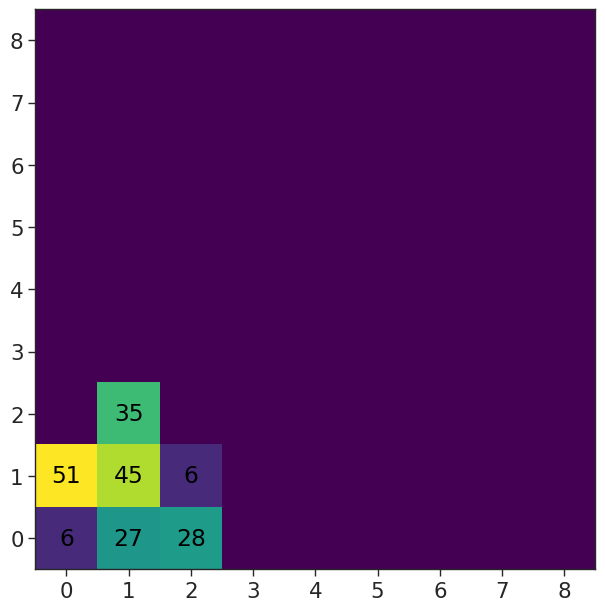}}\\
    \subfloat[70 $\le$ N. exp $<$ 90]{\includegraphics[width=.5\columnwidth]{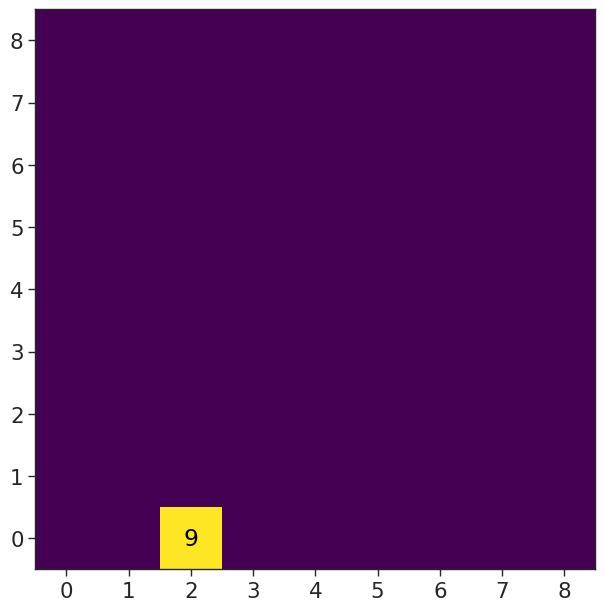}}\\
    \subfloat[N. exp $<$ 70]{\includegraphics[width=.5\columnwidth]{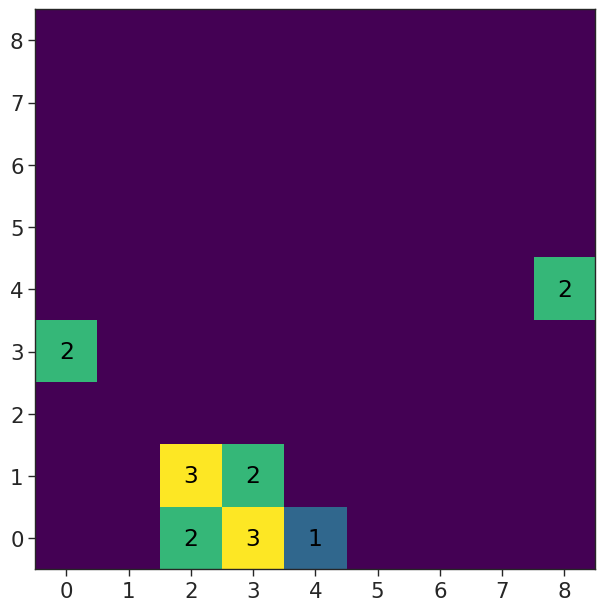}}

    \subfloat{\hspace{.2cm}\ViridisColorbar{0.5cm}}
    
    \caption{Distribution of the most recurrent Type 1 AGNs within the AGNs-majority cells of the map obtained with seed = 188. The color scale indicates the number of such objects associated with each neuron. \textit{Top panels}: Type 1 that populate AGNs-majority cells in more than 90 out of 100 experiments. \textit{Middle panel}: Type 1 that populate AGNs-majority cells in in 70 to 90 experiments. \textit{Bottom panel}: Type 1 that populate AGNs-majority cells in fewer than 70 experiments.}
    \label{fig:type1_freq}
\end{figure}

Specifically, the analysis focused on identifying those Type 1 and Type 2 AGNs that, across 100 experiments performed with different random seeds, frequently appear in cells with a majority AGNs population. The results show that a significant number of Type 1 AGNs (198 objects) populate AGNs-majority cells in more than 90 out of the 100 experiments. An additional 9 Type 1 appear in AGNs-majority cells in 70 to 90 experiments, while only 15 Type 1 AGNs are found in such cells in fewer than 70 experiments. Once these objects were identified, we examined which cells they populate in the experiment presented above using seed = 188. Figure~\ref{fig:type1_freq} maps their presence within the AGNs-majority cells, indicating how many of these objects populate each neuron. As can be seen, the AGNs-majority cells are mostly populated by the same Type 1 objects, while the more ``unstable'' ones are the objects that end up in cells contaminated by stars or galaxies.
It should be noted that the totals reported per cell do not necessarily correspond to the number of Type 1 objects shown in Figure~\ref{fig:red_top_som}, as these frequencies pertain exclusively to cells that exhibited a majority of AGNs in all the 100 experiments.

\begin{figure}[!htbp]
    \centering
    \subfloat[N. exp $\ge$ 90]{\includegraphics[width=.5\columnwidth]{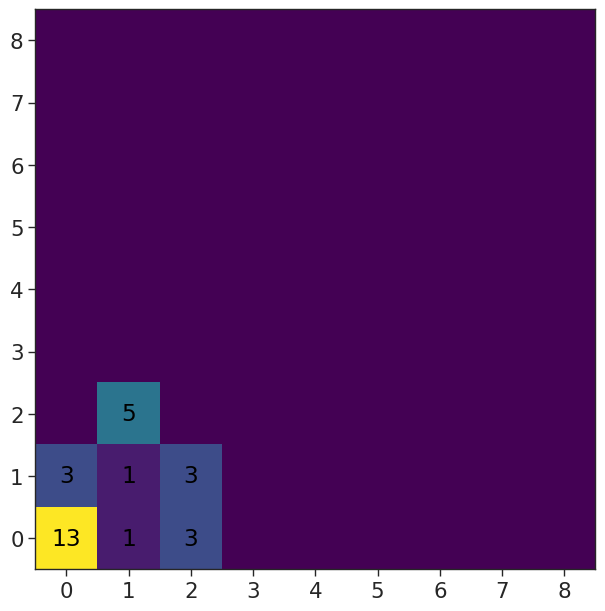}}\\
    \subfloat[N. exp $<$ 70]{\includegraphics[width=.5\columnwidth]{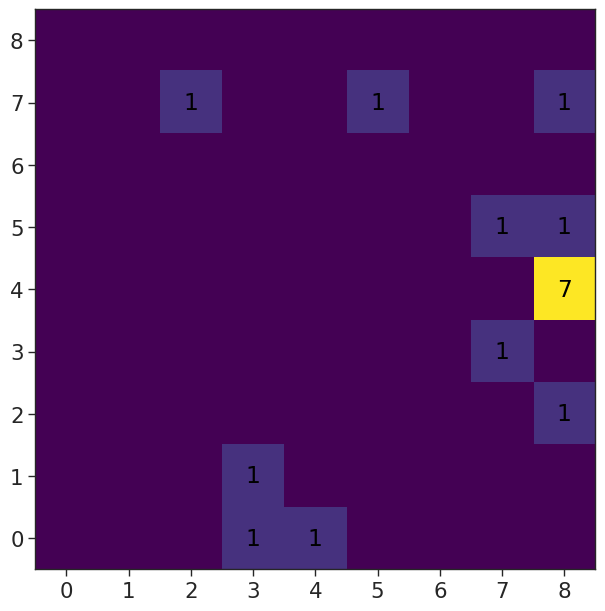}}

    \subfloat{\hspace{.2cm}\ViridisColorbar{0.5cm}}
    
    \caption{Distribution of the most recurrent Type 2 AGNs within the AGNs-majority cells of the map obtained with seed = 188. The color scale indicates the number of such objects associated with each neuron. \textit{Top panels}: Type 2 that populate AGNs-majority cells in more than 90 out of 100 experiments. \textit{Middle panel}: Type 2 that populate AGNs-majority cells in in 70 to 90 experiments. \textit{Bottom panel}: Type 2 that populate AGNs-majority cells in fewer than 70 experiments.
    }
    \label{fig:type2_freq}
\end{figure}

Similarly, we performed the same analysis for Type 2 AGNs. The results show that only 29 Type 2 AGNs populate AGNs-majority cells in more than 90 out of the 100 experiments, none appear in AGNs-majority cells between 70 and 90 experiments, and 17 Type 2 AGNs are found in such cells in fewer than 70 experiments. Then, we mapped their distribution across AGNs-majority cells in the experiment with seed = 188. The results, shown in Figure~\ref{fig:type2_freq}, reveal that 13 of the 29 objects cluster together in the unique ``uncontaminated'' neuron with a Type 2-majority (0,0), while 7 of the objects identified in fewer than 70 experiments are grouped in cell (8,4). Although cell (8,4) is a Type-2 majority neuron (see bottom panel of Figure~\ref{fig:red_top_som}), it is notably populated by 65 known galaxies and 526 unlabeled objects (see top panel of Figure~\ref{fig:red_top_som}). 

An analysis of the features for the 13 Type 2 and 6 Type 1 sources located in cell (0,0) reveals that their distributions are typically found in the tails, rather than near the peaks, of the corresponding LS distributions. Notably, they show particularly high values of both P$_{var}$ and $IAR_\phi$, which are indicative of strong and structured variability. Such behavior may suggest the presence of highly active or irregular processes, potentially associated with obscured or atypical AGNs activity.

Overall, the analysis highlights a clear difference in behavior between Type 1 and Type 2 within the AGNs-majority cells. Type 1 AGNs exhibit a highly consistent presence, with the vast majority populating AGNs-majority neurons across almost all experiments, suggesting a robust and well-defined clustering pattern. In contrast, Type 2 AGNs show a more scattered distribution with fewer objects consistently associated with AGNs-majority cells, reflecting more complex intrinsic differences in the feature-based representation between Type 1 and Type 2.

\subsubsection{Label propagation in the AGNs dominated cells}
Several cells predominantly containing AGNs also include unlabeled sources that exhibit AGN-like characteristics, particularly in their variability features. This pattern suggests that these unlabeled sources could potentially be AGNs, warranting further investigation.

As was previously mentioned, in a SOM, each cell corresponds to a neuron represented by a weight vector, whose length matches that of the input data (i.e., the number of features used for training). This weight vector acts as a prototype for all the objects for which the neuron is the BMU. To better characterize the AGNs-majority cells and strengthen the reliability of our analysis, it is useful to identify the object closest to the prototype for each neuron.
After determining the prototypes for all cells, we identified both the closest labeled and unlabeled object to each prototype. For AGNs-majority cells, we then verified whether the labeled object was itself classified as an AGN, and, for the unlabeled objects, we searched the Simbad database\footnote{\url{https://simbad.cds.unistra.fr/}} to determine if any of these sources had been independently identified as AGN.

\begin{table}[!htbp]
    \caption{Distribution of sources across AGNs-majority cells of the SOM.}
    \centering
     \resizebox{\columnwidth}{!}{
    \begin{tabular}{lcccccc}
        \hline
        \hline
        \textbf{Cell} & \multicolumn{3}{c}{\textbf{LS}} & \multicolumn{3}{c}{\textbf{US}} \\
        \textbf{[1]} & \textbf{[2]} & \textbf{[3] } & \textbf{[4]} & \textbf{[5]} & \textbf{[6]} & \textbf{[7]}\\
        \hline
        \rowcolor{lavender}
        (0,0) & 19 & (150.4183, 2.0851672)  & AGN$^{*}$ & 10 & (150.25323, 1.9661424) & AGN\\
        (1,0) & 28 & (150.20827, 1.8754041) & AGN$^{*}$ & 4 & (150.2556, 2.3534932) & AGN\\
        \rowcolor{lavender}
        (2,0) & 44 + 2 GAL& (149.77417, 2.6741626) & AGN & 49 & (149.79821, 2.3900828) & EmG$^{*}$\\
        (0,1) & 55 & (149.73895, 2.2206892) & AGN$^{*}$ & 5 & (149.77079, 1.7466647) & AGN\\
        \rowcolor{lavender}
        (1,1) & 46 & (150.45356, 2.5279411) & AGN$^{*}$ & 4 & (150.47956, 2.2531361) & AGN\\
        (2,1) & 13 + 2 STAR & (150.34245, 2.2262767) & AGN$^{*}$ & 9 & (150.53943, 1.9236345) & AGN\\
        \rowcolor{lavender}
        (1,2) & 42 & (150.44369, 2.0491065) & AGN$^{*}$ & 7 & (149.77293, 2.5557609) & SN/AGN\\
        (2,7) & 1 & (150.07368, 2.346843)  & AGN & 51 & (150.55828, 2.6438238) & GAL$^{*}$ \\
        (5,7) & 1 & (150.04032, 2.4712367) & AGN & 81 & (150.38622, 1.8416973) & GAL$^{*}$ \\

         \hline
    \end{tabular}
    }
    \tablefoot{Column \textbf{[1]} identifies the cell involved, Column \textbf{[2]} shows the number of source for which we know the classification from the Labeled Sources (LS): they are all AGN with the exception of cells (2,0) and (2,1), for which we report also the contaminants (galaxies or stars). In Column \textbf{[3]} we report the coordinates in the format (RA, Dec) of the closest labeled sources to the prototype of the given cell, and Column \textbf{[4]} shows how this source is labeled. Analogously, Column \textbf{[5]} reports the number of unlabeled sources falling into the given cell, Column \textbf{[6]} the coordinates of the closest unlabeled object to the prototype, and in Column \textbf{[7]} we report the classification (if available) from Simbad database; the classes are abbreviated in the following way: active galactic nucleus (AGN), emission-line galaxy (EmG), galaxy (GAL), supernova (SN). 
    Labels in Column [4] and [7] marked with an asterisk indicate whether the corresponding source is the closest to the prototype when considering both the LS and US.}
    \label{tab:som_prototype}
\end{table}

The results of this analysis are summarized in Table~\ref{tab:som_prototype}, which provides an overview of the distribution of both labeled and unlabeled sources across AGNs-majority cells. For each cell, we report the number of labeled objects, the coordinates and classification of the closest labeled source, as well as analogous information for the unlabeled population, including the SIMBAD classification. Sources marked with an asterisk indicate whether they are the overall closest to the prototype when considering both labeled and unlabeled data.

This detailed mapping allows us to validate the consistency of the SOM representation. In particular, the fact that the nearest labeled sources are predominantly AGNs strengthens our confidence in the ability of the map to capture meaningful structure in the feature space. Moreover, the identification of unlabeled sources with similar characteristics, and in some cases classified as confirmed AGNs in Simbad independent studies, opens the possibility for discovering new AGN candidates.

\section{Conclusions}\label{sec:conclusion}
Our analysis explored the effectiveness of using SOMs to classify AGNs within the COSMOS field, examining the roles of different feature sets based on variability, as well as the impact of including the Spitzer color (Ch21) as an additional feature. Once the best feature subset (\textit{Red - corr. + top Experiment}) have been defined through the study of four main indicator (Completeness Type 1, Completeness Type 2, Pureness AGNs, and Completeness AGNs), we ran the SOM with a fixed random seed and analyzed the results. Below are the primary insights and conclusions drawn from our experiments:

\paragraph{AGN completeness and Purity:} In the \textit{Red - corr. + top Experiment} with 100 random seeds, the SOM achieved a purity of $(97.5 \pm 1.6)\%$.
Completeness rates varied by AGNs type, with Type 1 objects showing a high completeness of $(93.2 \pm 2.6)\%$, while Type 2 AGNs were less complete $(25.0 \pm 1.4)\%$, consistent with previous findings about the difficulty of distinguishing Type 2 AGNs without extensive spectral data.

\paragraph{Role of Ch21:} While the addition of Ch21 provided subtle shifts in the results, it did not significantly alter the core results of all the experiments. This suggests that while Ch21 adds value, it does not distinctly impact AGNs classification when used in conjunction with other colors and variability features.

\paragraph{Anomalous cells:} Cells containing AGNs showed a different behavior with respect to the cells containing non AGNs in some key features, consistent with AGNs behavior. The SOM cells which show outliering behavior in some features often contained primarily AGNs, demonstrating that our variability features are effective indicators within the feature space. This also implies that cells with unlabeled sources but significant difference in terms of features could contain still unknown AGN candidates.

\paragraph{Unlabeled set:} Several cells with a majority of AGNs contained also unlabeled sources displaying hence AGN-like characteristics (e.g., similar variability features). This suggests they may indeed be AGNs, warranting further investigation. 
For such SOM cells we explored in literature, through the Simbad database, if the US sources closest to the prototypes of these cells have been flagged as AGN (see Table \ref{tab:som_prototype}). 
The results show that most of them are AGNs, which supports the reliability of the classification in these cells and suggests that also the other sources falling in those cells maybe worth to be considered at least as candidate AGNs.

\paragraph{Comparative results:} Compared to previous works \citep[See Table~5 of][]{decicco21}, our SOM-based classification method provides interesting results in purity, being able to achieve a result $\sim40\%$ better at the price of a decrease in terms of completeness of about $\sim12\%$, mostly due to a decrease in completeness of $\sim5\%$ of Type 2. 
This approach benefits from unsupervised learning’s ability to identify patterns without predefined labels, potentially enhancing AGNs detection rates and reducing contamination compared to traditional supervised methods, suffering also of the unbalancing of the data.

\paragraph{Summary:} The SOM-based approach shows significant promise for AGNs classification in large, complex datasets, especially through the use of variability features. Our results demonstrate a relatively efficient separation of Type 1 AGNs, stars, and galaxies, while the identification of the more elusive Type 2 remains a challenge, reflecting a known limitation in AGNs studies. Future work will focus on refining the feature set and expanding the unlabeled dataset, with the aim of improving the representation of the full AGNs population and enhancing classification performance in preparation for next-generation large-scale surveys such as LSST.

\begin{acknowledgements}
The authors thank the anonymous referee for the valuable comments and suggestions that have improved the quality of this manuscript. This paper is supported by Italian Research Center on High Performance Computing Big Data and Quantum Computing (ICSC), project funded by European Union - NextGenerationEU - and National Recovery and Resilience Plan (NRRP) - Mission 4 Component 2 within the activities of Spoke 3 (Astrophysics and Cosmos Observations).
DD acknowledges PON R\&I 2021, CUP E65F21002880003. DD and MP also acknowledge the financial contribution from PRIN-MIUR 2022 and from the Timedomes grant within the ``INAF 2023 Finanziamento della Ricerca Fondamentale''. MB, SC and GR acknoweldge the ASI-INAF TI agreement, 2018-23-HH.0 ``Attività scientifica per la missione Euclid - fase D''
SC and GR acknowledge support from PRIN MUR 2022 (20224MNC5A), ``Life, death and after-death of massive stars'', funded by European Union – Next Generation EU.
Topcat \citep{Taylor05} and STILTS \citep{Taylor06} have been used for this work.
Some of the resources from \cite{Stutz2022} has been used for this work.
Some of the methods used in this work are part of the Scikit package \citep{Pedregosa11}.
The SOM used in this work is part of the python package MiniSOM \citep{vettigliminisom}.

\end{acknowledgements}

\bibliographystyle{aa}
\bibliography{main} 

\end{document}